\documentclass[final]{svjour2}
\usepackage{graphicx}
\usepackage{rotating}
\usepackage{amssymb}
\usepackage{mathptmx}
\usepackage[numbers]{natbib}
\makeatletter
\journalname{Journal of Low Temperature Physics}

\bibpunct{}{}{,}{s}{}{,}

\begin{document}

\newcommand{\hdblarrow}{H\makebox[0.9ex][l]{$\downdownarrows$}-}
\title{Silicon Vibrating Wires at Low Temperatures}

\author{Eddy Collin$^{*}$ \and Laure Filleau \and Thierry Fournier \and Yuriy M. Bunkov \and Henri Godfrin}

\institute{Institut N\'eel / CNRS and Universit\'e Joseph Fourier - 25, rue des Martyrs  -  Bt. E - BP 166 \\ 38042 Grenoble cedex 9 – France \\
\email{eddy.collin@grenoble.cnrs.fr}}

\date{14.12.2007}

\maketitle


\begin{abstract}
Nowadays microfabrication techniques originating from micro-electro\-nics enable to create mechanical objects of micron-size. The field of Micro-Electro-Mechanical devices (MEMs) is continuously expanding, with an amazingly broad range of applications at room temperature.  \\
Vibrating objects (torsional oscillators, vibrating wires) widely used at low temperatures to study quantum fluids, can be replaced advantageously by Silicon MEMs. In this letter we report on the study of Silicon vibrating wire devices. A goal-post structure covered with a metal layer is driven at resonance by the Laplace force acting on a current in a magnetic field, while the induced voltage arising from the cut magnetic flux allows to detect the motion. The characteristics of the resonance have been studied from 10~mK to 30~K, in vacuum and in $^4$He gas. In this article, we focus on the results obtained above 1.5~K, in vacuum and gas, and introduce some features observed at lower temperatures. \\
The resonant properties can be quantitatively understood by means of simple models, from the linear regime to a highly non-linear response at strong drives. We demonstrate that the non-linearity is mostly due to the geometry of the vibrators. We also show that in our device the friction mechanisms originate in the metallic layers, and can be fully characterized. The interaction with $^4$He gas is fit to theory without adjustable parameters. \\


\end{abstract}


\section{INTRODUCTION}

Vibrating wires are standard probes for $^3$He and $^4$He liquids in their normal and superfluid states\cite{viscosity1,viscosity2,pickett}. 
With the advent of microfabrication, micro-electro-mechanical (MEM) structures can be made and optimized\cite{MEMsBOOKS} in order to replace and transcend this common technique\cite{heliumMEMs}.  \\
Actually, Silicon vibrating wires are an essential feature of the ULTIMA project, in which the unique properties of superfluid $^3$He are used to make ultra-sensitive bolometers for cosmic particle detection\cite{ultima}. \\
Several studies on Silicon torsional oscillators at low temperatures have been published\cite{kleimanbishop,jeevak}. In the most general context, many applications of low temperature probes and actuators based on torsion or flexion of Silicon structures can be envisaged. The understanding of the mechanical properties of these devices at low temperatures is thus a  pre-requisite.

\vspace{0.15in}
In this paper we present our results on micron-size goal-post shaped Silicon structures. 
After a brief introduction of the devices, we describe our experimental setup and the signatures observed in the measurements. A third theoretical part develops the calculations necessary for the quantitative data analysis of the fourth part.
The latter part is a brief introduction to irregularities due to very strong drives, and effects observed at the lowest temperatures. The conclusion summarizes our understanding of these devices, and introduces some future work.  
 
\subsection{The structure}

\begin{figure}
\vspace*{2. cm}
\centerline{\includegraphics[height = 14.5 cm]{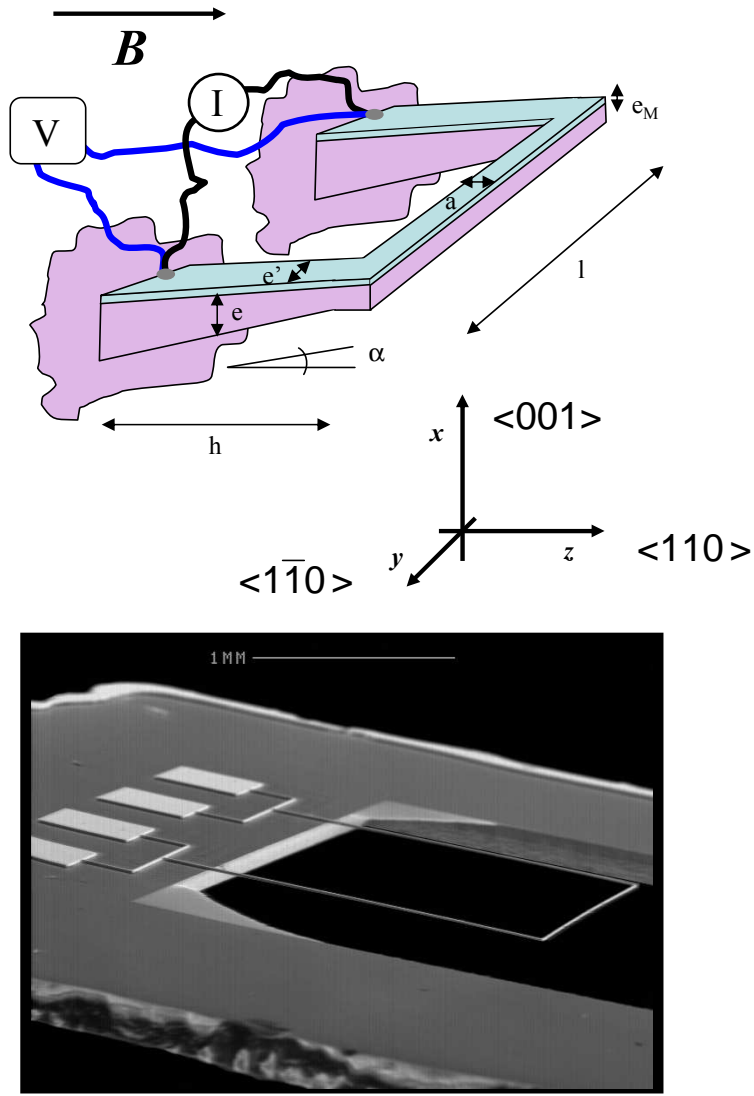}}
%
\caption{(Color online). Top: schematic of the samples studied. The driving method is represented with $\vec{B}$ and $I$, the detection with $V$. The darker color is Silicon while the lighter color on top is metal. The geometrical parameters are presented, together with the axis and their corresponding crystallographic directions. Bottom: actual SEM picture of one sample (Eb2), with contact pads. The sample thickness is about 10$\,\mu$m.}  
\label{fig1}
\end{figure}

The mechanical structure we consider is shown on Fig. \ref{fig1}. The fabrication method has already been reported\cite{usPhysicaB}, but we shall quickly recall it. The sample is made from (low Boron-doped) monocrystalline $<\!\!001\!\!>$ Silicon wafers (typically resistivity $\rho \approx$ 25 m$\Omega.$cm at 300$\,$K) protected by Si$_{3}$N$_{4}$. We start by chemically etching (KOH) a window on the back of the sample, leaving only a thin membrane. On the front side, a metallic layer is deposited either by Joule evaporation or by means of a magnetron. A masking layer having the desired goal-post shape is then patterned, and protects an area of the sample from the last RIE step (Reactive Ion Etching). With an Aluminum metallic layer, we use the metal itself as a mask for the Silicon. Otherwise, resist is used and removed by an $O_2$ plasma. In the last etching step, the membrane is etched away releasing the vibrating structure. A SEM (Scanning Electron Microscope) picture of a sample is presented in Fig. \ref{fig1}, bottom.

\vspace{0.15in}
The simplified theoretical structure we consider consists of two (identical) rectangular cantilever beams (called thereafter "feet") of length $h$ and width $e'$ (Fig. \ref{fig1}, top). They are linked by a "paddle" of length $l$ and width $a$. The average Silicon thickness is $e$, while the feet have an overall thickness gradient $\alpha=\frac{de}{dz}=\,$Cst (due to the chemical etching method). \\
On top of the structure a metallic layer of thickness $e_{M}$ ensures the drive and detection. The Laplace force $\int{\! I dl \, \vec{y} \wedge B \, \vec{z}}= I l B \, \vec{x}$ acting on the paddle generates the distortion in the $\vec{x}$ direction, and the motion induces a voltage $V= d \phi_{cut}/dt = B l\, d X_m/\!dt \, \vec{z}.\vec{\mathsf{z_m}}$ (Lenz's law). $d X_m/\!dt$ is the speed of the extremity ($\vec{\mathsf{z_m}}$ is the direction along the distorted feet, at their end point). The effects of the metal on the mechanical properties of the structure are discussed in secion \ref{themetal}.

\section{EXPERIMENT}
\label{experiment}

\begin{figure}
\centerline{\includegraphics[width = 9. cm]{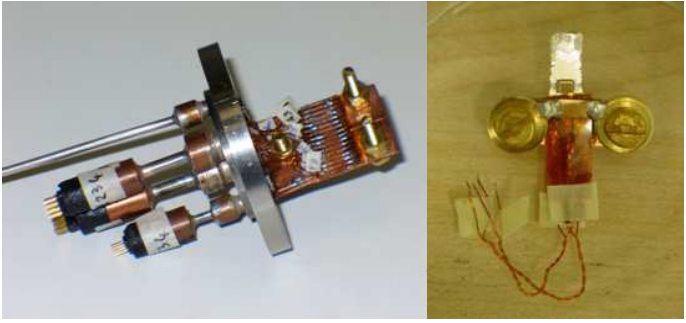}}
%
\caption{(Color online). Left: Copper support receiving the sample-holders with connecting stripes (mounted on the top-hat of the cell). On one side a thermometer while on the other a heater are mounted. Right : A sample (E6) clipped in a sample-holder (beryllium-copper springs).}  
\label{cell}
\end{figure}

The results reported in this article have been obtained in a $^4$He pumped cryostat, down to 1.5$\,$K. The Silicon samples were mounted on a thick (1$\,$mm) copper plate in a pumped cell (Fig. \ref{cell}). On one side of the copper plate, a calibrated Allen-Bradley carbon resistor has been affixed, while on the other side we mounted a 100$\,\Omega$ heater. The drive/detection setup is based on the standard vibrating wire scheme\cite{CHH}. A twisted pair of wires allows to drive the sample with a current $I$, while another twisted pair allows us to detect the induced voltage $V$. A copper coil (diameter 35$\,$mm, length 61$\,$mm) generating  the external field $B$ (27.58$\,$mT/A at its center) surrounds the cell. The inhomogeneity over a couple of mm in the center has been calculated to be smaller than 1.5$\,\%$.\\
The signal $V$ is brought to a (high impedance) room temperature differential preamplifier, and fed to a lock-in detector. A DC current source provides the current for the coil, while an arbitrary waveform generator applying a voltage onto a 100$\,$k$\Omega$/10$\,$k$\Omega$/1$\,$k$\Omega$ resistor (in series with the low impedance Silicon vibrating wire) generates the current $I$. A commercial low frequency telecoms $1:1$ transformer has been used to decouple the drive from the detection. The setup has been carefully calibrated in order to provide an absolute error on $B$, $I$ and $V$ of the order $\sim\,1\,$\% each. However, alignment and centering of the Silicon sample with respect to the coil are certainly imperfect, and could be the source of additional errors. The resolution on $V$ depends obviously on the signal strength, which varied from 0.1$\,$mV down to a fraction of $\mu\!$V. \\
A diffusion pump produces (at the high temperature end of the pumping line) a vacuum of the order of $5. \,10^{-6}\,$Torr. We verified that continuously pumping on the cell, or closing the room temperature valve to the cell after a night's pumping and a cool down to 4.2$\,$K gave the same experimental results. We thus call this low pressure limit our {\em vacuum limit}. Measurements in $^4$He gas at 4.2$\,$K are presented in section \ref{4he}. \\
The temperature was measured and regulated with a resistance bridge. Only the copper plate is regulated, the stainless steel cell is kept immersed in liquid Helium. For each change of temperature, a settling time of at least half an hour was allowed. The calibration of the thermometer is believed to be accurate to $\sim\,2\,$\% approximately.

\begin{figure}
%
\centerline{\includegraphics[width= 9. cm]{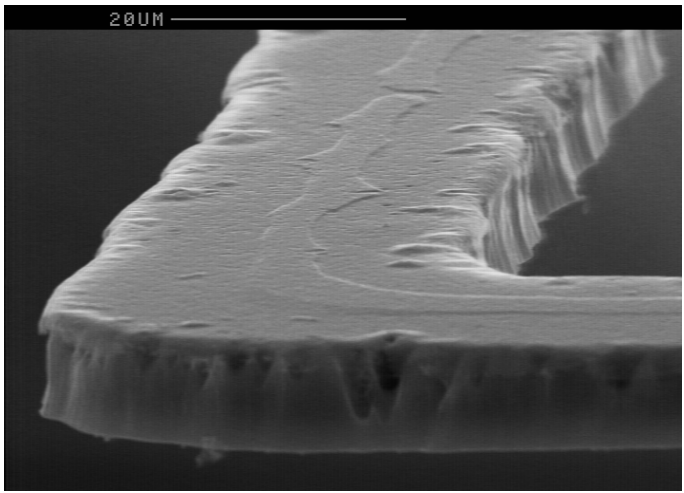}}
%
\caption{SEM close-up of the left corner of sample E6 with its first Aluminum deposition (on top). Etching structures are visible on the metallic layer, which slightly overruns the Silicon. The poor straightness of the Silicon shape is due to the rather poor quality of the optical lithography.}  
\label{Allayer}
\end{figure}

\vspace{0.15in}
In the following we present the measured resonance lines, and infer experimentally the coefficients and parameters describing the vibrating structure and its constitutive materials. Various samples have been tested, and on one of them (E6) we added metal {\em two times, on both sides}, in order to properly extract its contribution. 
We used the Joule evaporation technique. The metal (99.99$\,$\%-Al) was deposited at a rate comprised between 10$\,$\r{A}/s and 40$\,$\r{A}/s, in a vacuum of about $10^{-5}\,$Torr. The sample was {\em not} cleaned between evaporations. The first metallic deposition served also as a protection for the Silicon structure in the last etching step (RIE) of the fabrication process. A close-up SEM image is shown in Fig. \ref{Allayer}. \\
 The theoretical expressions of part \ref{theor} are used in the quantitative data analysis which is postponed to part \ref{datanalys}. These theoretical fits and the careful measurements realized in a very broad range of parameters at low temperatures, constitute the core of this article, demonstrating our understanding of these devices. 

\subsection{Resonance properties}

\begin{figure}
\centerline{\includegraphics[width= 12. cm]{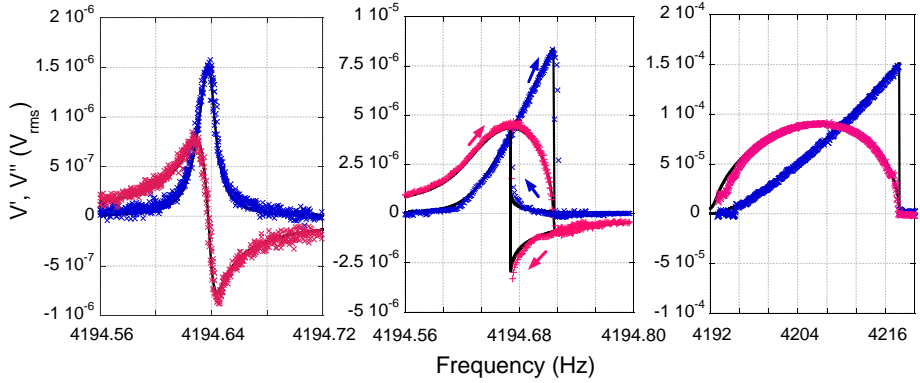}}
%
\caption{(Color online). Three resonance lines measured in vacuum at 4.2$\,$K with different drives (sample E6 with first metal deposition). The magnetic field was 11$\,$mT. The dark (or blue) crosses are the in-phase signals $V'$, while the light (pink) crosses are the out-of-phase $V''$ voltages. From left to right: linear signal, force 11$\,$pN$_{rms}$ and displacement 3.3$\,\mu$m$_{rms}$; non-linear signal, force 69$\,$pN$_{rms}$ and displacement 18$\,\mu$m$_{rms}$ (the arrows indicate the sweep direction, showing hysteresis); very non-linear signal, force 4.6$\,$nN$_{rms}$ and displacement 0.33$\,$mm$_{rms}$ (up sweep). The black lines are fits (see text).}  
\label{3lines}
\end{figure}

In Fig. \ref{3lines}, we present three typical resonance lines obtained for very different drives in vacuum at 4.2$\,$K. We define, written in complex terms, $\tilde{V}(\omega)= \,V' + i\, V''$ for the detected voltage $V$ in Fourier space, with $\omega=2 \pi f$ the angular frequency (and $i=\sqrt{-1}$). 
In part \ref{theor}, the (harmonic) displacement $X_m(t)$ of the end of the structure is described with the resonance components $X\!P_m$ and $X\!Q_m$ (in-phase and out-of-phase displacements respectively). 
$V' = l B \,\omega \,X\!Q_m$ is the in-phase signal detected, and $V'' = l B \,\omega \,X\!P_m$ the out-of-phase signal.
$V'$ is a peaked function and from its maximum $V'_{max}$ we calculate the maximum displacement $X_{max}$ reached by the top of the structure. Most of the values quoted in this article are {\it rms} (root-mean-square), which derive from the peak amplitude with $x_{rms}=x_0/\sqrt{2}$. 
The signal is measured while sweeping the frequency (usually upwards otherwise the direction is stipulated). For each point, a settling time of the order of $1/(\pi \,\Delta f)$ has been taken to reach the steady-state limit ($\Delta f$ being the full width at half height of the $V'$ component). \\
The resonance lines are studied for various temperatures and driving forces. At small drives, the response of the mechanical system is always linear, while for higher drives we are able to reach a highly non-linear regime. This is presented in Fig. \ref{3lines}.
From left to right, we show a resonance obtained at the end of the linear regime, then a non-linear measurement displaying the usual features of non-linear resonances, and at last a very non-linear line obtained for displacements $X_{max}$ of the order of $0.33\,$mm$_{rms}$. This last measurement is remarkable, in the sense that the peak-to-peak displacement is of the order of 
$0.9\,$mm for a structure of length $h=1.35\,$mm. The distortion is {\em extreme}, the angle swept by the structure being of the order of 40$^\circ$. 

\begin{figure}
\centerline{\includegraphics[width= 9 cm]{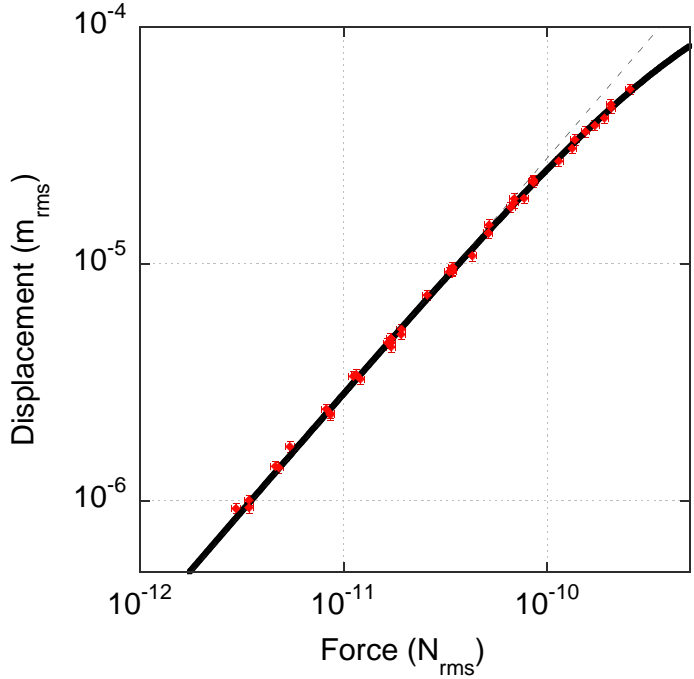}}
%
\caption{(Color online). Displacement $X_{max}$ versus force $F_{ac}$ curve for sample E6 (first metal deposition) in vacuum at 4.2$\,$K (logarithmic scale). The thick line is a fit, which associated to the measurement of the full width $\Delta f$ enables to extract the vibrating mass of the object (Tab. \ref{NonlinSamples}). At strong drives, non-linear effects are already visible (discrepancy between dashed and full lines) and will be discussed in the following paragraphs. Error bars are of the order of 5$\,$\% altogehter. }  
\label{mass}
\end{figure}

\vspace{0.15in}
The black lines are fits, with $X\!P_m$ and $X\!Q_m$ obtained from Eq. (\ref{nonlinphase}) and Eq. (\ref{nonlinquad}) respectively (part \ref{theor}). For small displacements (small drives), they reduce to the {\em Lorentz} expressions, Eq. (\ref{phase}) and Eq. (\ref{quad}). \\
In this linear regime (left graph Fig. \ref{3lines}), some useful and well-known properties can be given for $\tilde{V}(\omega)$. The $V'$ component is maximum at $f=f_{res}$, with $V'' =0$ (resonance point). The height $V'_{max}$ is proportional to the force $F_{ac}$ applied, and inversely proportional to the full-width-at-half-height $\Delta f$ of the $V'$ line. In fact, this property can be used to "weigh" experimentally the wire, extracting the vibrating (or normal) mass $m_{v0}$ (see Fig. \ref{mass}). For sample E6 and the first metal deposition, we obtained $1.35\,\mu$g (Tab. \ref{NonlinSamples}). The quality factor defined as $Q=f_{res}/\Delta f$ is of the order of $0.3\,10^{6}$ which is a remarkable figure. \\
When the driving force becomes too strong, the resonance enters the non-linear regime. One has then to solve numerically (see fits of  Fig. \ref{3lines}) Eq. (\ref{nonlinphase}) and Eq. (\ref{nonlinquad}). At first, the resonance line is pulled toward higher frequencies (for a positive non-linear coefficient $b_{0}$ introduced in part \ref{theor}; it is obviously inverted for the other polarity). At some critical drive, the equations to solve get multi-valued: two stable solutions appear, one stable while sweeping the frequency up, and the other while sweeping down (see middle graph Fig. \ref{3lines}). Hysteresis appears, and close to the maximal amplitude point on the $V'$ component, the device switches from motion to stillness (so-called bifurcation point). The in-phase signal $V'$ looks practically triangular, and the out-of-phase $V''$ practically circular. The last remains almost on the positive side, even above the bifurcation point. At very strong drives, the hysteresis is so large that practically no resonance can be detected on the down sweeps (right graph Fig. \ref{3lines}). \\
However, some simple criteria still do apply. If one sweeps the frequency upwards (in fact, in the direction corresponding to the polarity of the non-linear coefficient $b_{0}$), the signal $V'$ is still maximum at the so-called resonance frequency $f_{res}$ defined in section \ref{landau}, and $V''$ is still zero. The maximum height $V'_{max}$ of the peak $V'$ is still proportional to $F_{ac}$ and inversely proportional to $\Delta f$. But now, $f_{res}$ and $\Delta f$ are {\em functions of the displacement} $X_{max}$ (involving the coefficients $b_0$ and $b_1$ presented in part \ref{theor}), which in turn has to be evaluated at resonance. 
As a consequence, $\Delta f$ cannot be measured anymore as the half-height width of the $V'$ component. \\
Moreover, the theoretical model of section \ref{landau} predicts a non-linear detected signal $V$, through the $b_3$ parameter, which implies a non-linear dependence of the maximum measured height $V'_{max}$ with respect to $X_{max}$. 
The key result of the theoretical description is that {\em if the non-linearities originate from the geometry of the distortion}, all non-linear parameters ($f_{res}$, $\Delta f$, $V'_{max}$)
have to display a $(X_{max})^2$ dependency, from which the corresponding $b_i$ factor can be experimentally determined.
\begin{figure}
\centerline{\includegraphics[width= 9 cm]{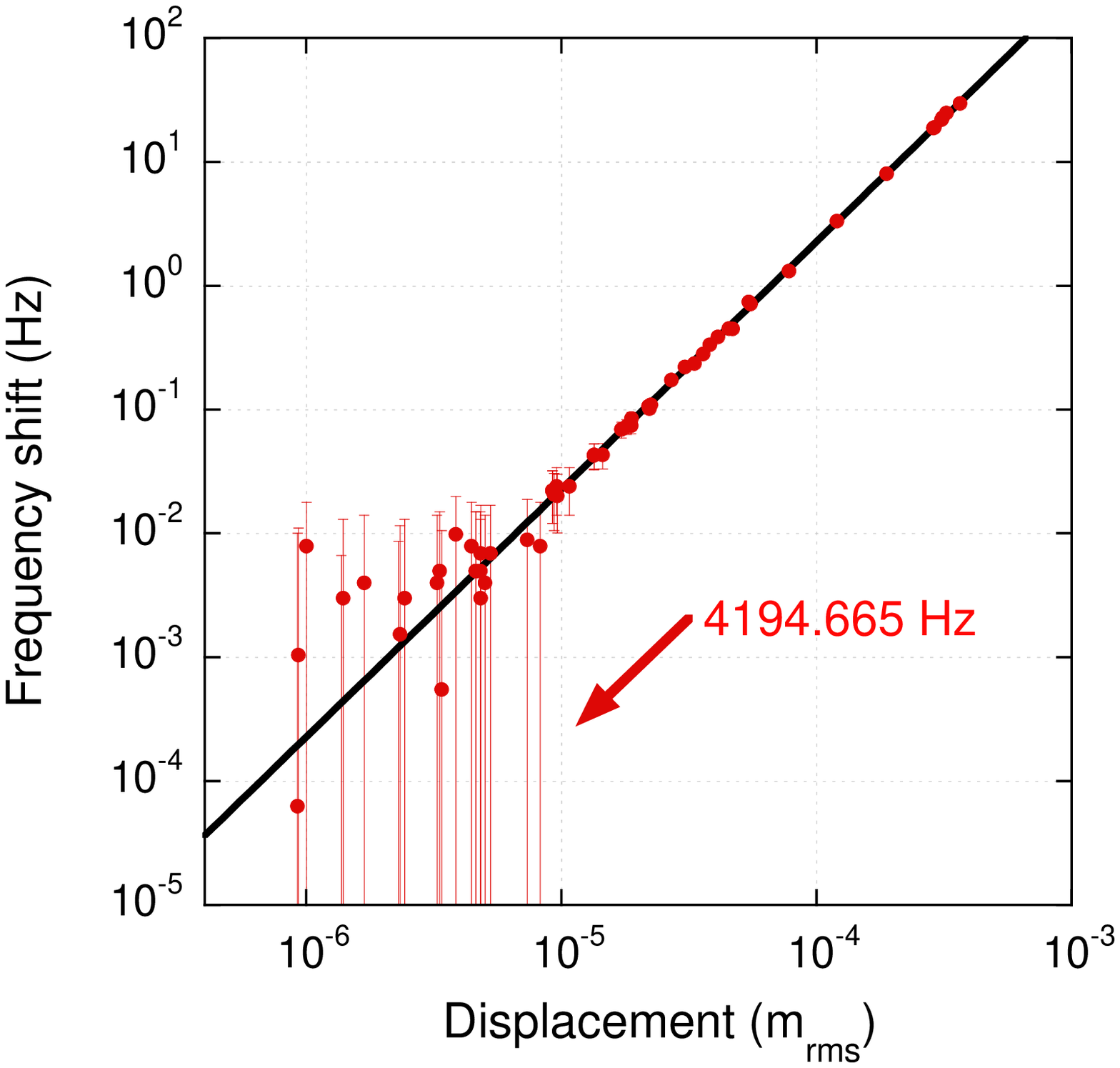}}
%
\caption{(Color online). Non-linear frequency shift as a function of the displacement $X_{max}$ (note the logarithmic scale), measured in vacuum at 4.2$\,$K (sample E6, first metal deposition). The linear resonance frequency given on the picture has been subtracted.  The line is the quadratic dependence expected from theory, fit on the data. }  
\label{freqnlin}
\end{figure}

\vspace{0.15in}
The non-linear behavior of the resonance frequency is presented on Fig. \ref{freqnlin}. Indeed, the dependence of the frequency shift on the displacement is quadratic, over the entire explored range, {\em including} various magnetic fields and various (non-linear) dampings. 
On the other hand, Fig. \ref{Kfact} presents the inverse normalized height $1/(V'_{max}/F_{ac})$. This parameter depends directly on $X_{max}$ through the $b_3$ factor, but is also simply proportional to the linewidth $\Delta f$ (damping term).
The striking result of Fig. \ref{Kfact} is that this parameter appears clearly to be a {\em linear function} of the displacement, in complete disagreement with the (simple geometrical) theory predicting a quadratic dependence.  \\
The apparent paradox is lifted as soon as we realize that the increase of the dissipation (increase in $\Delta f$) is due to the materials' response rather than simple (geometrical) non-linearities.
 In order to unambiguously prove that the linear dependence to $X_{max}$ of the inverse normalized height  originates in $\Delta f$, one can consider the {\em power balance} at resonance.
Indeed, exactly at resonance (where drive $F_{ac}$ and speed $\dot{x}$ are in-phase, since $V'' =0$), the (mechanical) potential energy $1/2 \,k_{v0} \, x^2$ is simply transferred to kinetic energy $1/2 \,m_{v0} \, \dot{x}^2$ and vice-versa (through $k_{v0} \, x\, \dot{x} = - m_{v0} \, \dot{x} \, \ddot{x}$), while the drive balances the dissipated energy. This means that the (electrical) injected power $F_{ac} \, \dot{x}= V\, I$ is exactly equal to $\Delta \omega \, m_{v0}\, \dot{x}^2$ (the dissipated power). The key point is that the calculated injected power depends {\em linearly on the detected voltage and is independent of the measured width}, while the calculated dissipated power depends {\em quadratically on the detected voltage and is proportional to the measured width}. Assuming that the non-linear height (Fig. \ref{Kfact}) is indeed a direct image of the width, we plot on Fig. \ref{powerbalance} the powers averaged over one oscillation period. The calculated points are exactly on the equality line (full black line) which proves that the assumption is correct: the linewidth $\Delta f$ is a {\em linear function of displacement}, or speed.

\vspace{0.15in}
We thus have to conclude that quadratic (geometrical) non-linear corrections on the resonance's height and width are negligible, a point explained in part \ref{datanalys}. \\
Furthermore, as in any low temperature experiment, it is legitimate to consider thermal gradients and question the temperature's homogeneity. Indeed, the line\-width's linear dependence to the displacement could be obtained with a thermal model. However, a quantitative analysis described in the following section allows us to rule out this possibility for our devices. 

\subsection{Absence of thermal gradients}
\label{thermgrad}

Two sources of heating are present, the mechanical power dissipated $\Delta \omega \, m_{v0}\, \dot{x}^2$ already introduced, and the Joule heating due to the resistivity of the metal layer $R_{el} \, I^2$. We write $I = I_0 \, \cos(\omega t)$ and $\dot{x}=\dot{x}_0 \,\cos(\omega t + \varphi) $.
For the sake of simplicity, we consider that this power is created at the end of the structure, and leaks toward the thermal bath (the copper support) through the metal layer. As shall be discussed in part \ref{datanalys}, since the dissipation process takes place in the metal itself, we basically neglect completely the Silicon$^($\footnote{A more complete model would only give a better thermal link; distributing the power dissipated along the feet of the structure reduces the temperature gradients too.}$^)$. \\
Disregarding non-linear effects, we write for the heat equation:
\begin{displaymath}
\rho_M C_M \frac{\partial T}{\partial t} - \kappa_M \frac{\partial^2 T}{\partial z^2} = 0
\end{displaymath}
with $\rho_M$, $C_M$ and $\kappa_M$ the density, specific heat and thermal conductivity of the metal, respectively. The boundary conditions are:
\begin{eqnarray*}
T(z=0,t) & = & T_0 , \\
\kappa_M \, e' e_M \, \frac{\partial T}{\partial z} (z=h,t) & = &  \frac{1}{2} \dot{Q}_0 +  \frac{1}{2} \Delta \dot{Q} \, \cos(2 \omega t + \phi)
\end{eqnarray*}
where the first equation stands for the bath temperature, and the last for the heat flow (the factor $ \frac{1}{2}$ reminding that there are two feet in the structure). With our notations:
\begin{eqnarray*}
\dot{Q}_0 & = &  \frac{1}{2} \Delta \omega \, m_{v0}\, \dot{x}_0^2\,\left(1  + R_{el} \frac{m_{v0} \, \Delta \omega}{(l\,B)^2}   \right) ,\\
\Delta \dot{Q} & = &  \frac{1}{2} \Delta \omega \, m_{v0}\,\dot{x}_0^2 \sqrt{1+ R_{el} \frac{m_{v0} \, \Delta \omega}{(l\,B)^2} \left(2 \cos(2 \varphi) + R_{el} \frac{m_{v0} \, \Delta \omega}{(l\,B)^2} \right)}
\end{eqnarray*}
while $\tan \phi=\sin(2 \varphi) / [ R_{el} \frac{m_{v0} \, \Delta \omega}{(l\,B)^2} + \cos(2 \varphi)]$. We define $C_{th}=\rho_M C_M (h \,e' e_M)$ the heat capacity and $R_{th}= \kappa_M^{-1} h /(e' e_M)$ the thermal resistance of one foot of the structure. $\tau= R_{th} C_{th}$ is the thermalization time constant. The temperature at the end of one foot is:
\begin{eqnarray}
T(z=h,t) & = & T_0 + \frac{1}{2}\frac{\dot{Q}_0 h}{\kappa_M \, e' e_M } + \frac{1}{2}\frac{\Delta \dot{Q} h}{\kappa_M \, e' e_M }\, \frac{1}{\sqrt{\omega \tau }} \cos(2 \omega t + \phi-\pi/4), \label{poorT} \\
 & or &   \nonumber \\
T(z=h,t) & = &  T_0 + \frac{1}{2}\frac{\dot{Q}_0 h}{\kappa_M \, e' e_M } + \frac{1}{2}\frac{\Delta \dot{Q} h}{\kappa_M \, e' e_M }\, \frac{1}{\sqrt{2}} \cos(2 \omega t + \phi+\pi/4)
\end{eqnarray}
where the first equation stands for a poor heat conduction ($\omega \tau  \gg 1$), and the last for a good heat conduction ($\omega \tau  \ll 1$). 

\begin{figure}
\centerline{\includegraphics[width= 9 cm]{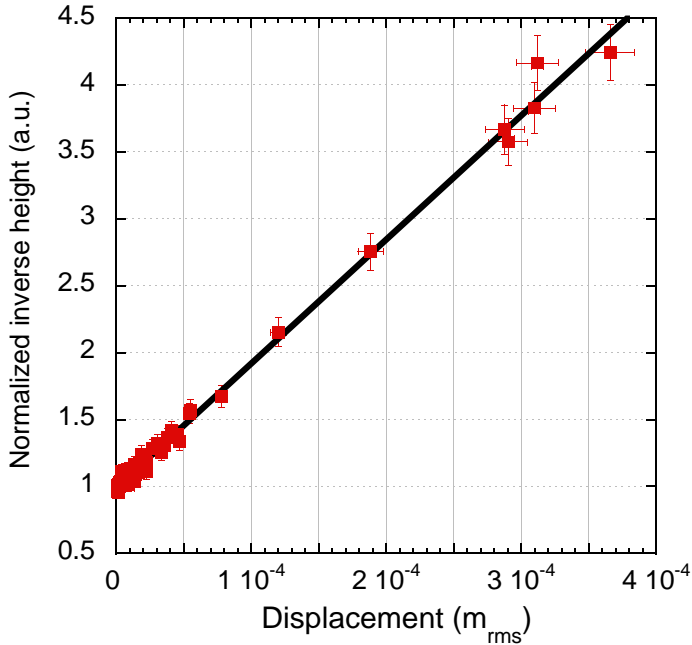}}
%
\caption{(Color online). Inverse normalized height $1/(V'_{max}/F_{ac})$ measured in vacuum at 4.2$\,$K for sample E6 (first metal deposition), as a function of displacement $X_{max}$. The line is a linear dependence, fit on the data, which is {\em not expected} from the geometrical theory of non-linearities.}  
\label{Kfact}
\end{figure}

\begin{figure}
\centerline{\includegraphics[width= 9 cm]{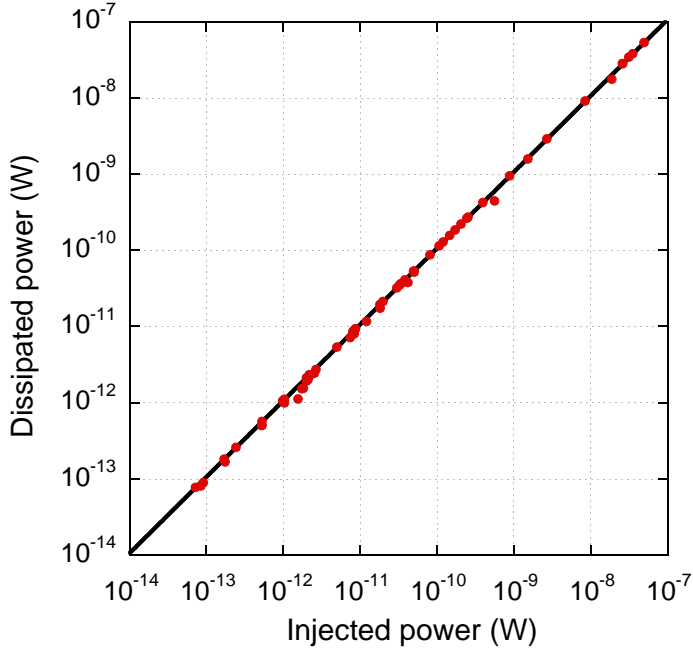}}
%
\caption{(Color online). Power balance obtained from Fig. \ref{Kfact}, assuming that the height non-linearity arises solely from the damping term $\Delta f$. The line represents equality of powers injected and dissipated (see text). Symbols are as large as $\pm\,$5\% error bars.}  
\label{powerbalance}
\end{figure}

\vspace{0.3in}
We take for the measured damping parameter $\Delta \omega = 2 \pi \Delta f$ a linear function of temperature $T$, as obtained experimentally in part \ref{datanalys}, producing $\Delta \omega(T)=\Delta \omega(T_0) + \beta \,\Delta T$. $\Delta T$ is the temperature increase of the region where friction occurs; in our simplified model the end of one foot. $\beta$ is an experimental parameter characterizing the intrinsic damping mechanism, which has to be measured. \\
We take for the metal's thermal conductivity $\kappa_M= \kappa_0 \, T$ and for the electrical resistivity a constant value $\rho_{el}$. Both are linked through the Wiedemann-Franz law $\kappa_M\,\rho_{el}=L \,T$ with $L$ the Lorenz number (see Ref. \cite{alu} for a discussion on Aluminum). The specific heat of the metal layer\cite{alu2} writes $C_M = \gamma \, T$. We express the (total) electric resistance as $R_{el} =\rho_{el} \left[ (2 h)/(e_M \, e')+  l/(e_M \, a)\right]$. \\
Two cases have to be distinguished:
\begin{itemize}
\item Poor thermal link: the temperature's oscillatory part in Eq. (\ref{poorT}) averages out, and only a (large) static temperature increase is present at the end of the structure. It writes at the lowest order:
\begin{displaymath}
\!\!\!\!\Delta T = \dot{x}_{rms} \, \sqrt{\frac{1}{2} m_{v0} \, \Delta \omega(T_0)\, \left(1+ \frac{R_{el}\,m_{v0} \, \Delta \omega(T_0)}{(l\,B)^2} \right)\,R_{th}(T_0) \,T_0} - \frac{1}{2} T_0  
\end{displaymath}
with $\dot{x}_{rms}=\dot{x}_0/\sqrt{2}$. The linear dependence to $\dot{x}_{rms}$ generates in turn a linear dependence to $\dot{x}_{rms}$ in $\Delta \omega$. Although it reproduces the {\em analytic} dependence measured (Fig. \ref{Kfact}), a {\em quantitative} analysis fails$^($\footnote{A fit of the data using this expression generates for the Aluminum metallic layer such a bad thermal conductivity that the RRR (so-called Residual Resistivity Ratio) would be {\em smaller} than 1, which is unphysical.}$^)$. Our devices are {\em never} in this limit.
\item Good thermal link: the temperature at the end of the structure oscillates with an amplitude almost as large as the average thermal gradient. However, whether the dissipation process will follow this oscillation or not depends on its intrinsic characteristic timescale, which is unknown. The (small) static temperature gradient writes at first order:
\begin{displaymath}
\!\!\!\!\Delta T = \dot{x}_{rms}^2 \, \frac{1}{2} m_{v0} \, \Delta \omega(T_0)\, \left(1+ \frac{R_{el}\,m_{v0} \, \Delta \omega(T_0)}{(l\,B)^2} \right)\,R_{th}(T_0)  
\end{displaymath}
which displays a quadratic dependence to $\dot{x}_{rms}$. Using values from the literature for Aluminum\cite{alu,alu2,alusupra} demonstrates that $\Delta T \ll T_0$, and the thermal contact is always excellent$^($\footnote{The discussion basically holds for any non-superconducting metal with a reasonable RRR (Residual Resistivity Ratio), with any (reasonable) thickness, and not too low temperatures.}$^)$: the temperature of the vibrating structure is uniform to an excellent approximation, and always equal to the measured temperature $T_0$.
\end{itemize}
We conclude that the linear dependence of the resonance curve's inverse height as a function of the displacement, seen in Fig. \ref{Kfact}, is genuine and originates from an {\em intrinsic} linear dependence of  $\Delta f$ to $X_{max}$. 
This effect is clearly due to the materials, and shall be described by some specific model: we introduce non-linear damping coefficients $\Lambda_{1} '$ and $\Lambda_{2} '$ arising from the {\em shear thickening fluid model} presented in section \ref{dissipe}.


\subsection{Fits of the resonance curves}

The black lines in Fig \ref{3lines} are fits obtained by solving numerically Eq. (\ref{nonlinphase}) and Eq. (\ref{nonlinquad}).
To obtain these fits, {\em only one set of parameters} is used over the entire parameter space (drive, temperature, fields, and even pressure). These parameters are in agreement with direct measurements like the determination of the mass (Fig. \ref{mass}), or the (geometrical) non-linear $b_0$ parameter and the (friction) non-linear $\Lambda_{1,2} '$ parameters (Figs. \ref{freqnlin} and \ref{Kfact} respectively). The values extracted from these analyses shall be discussed in part \ref{datanalys}. \\
The data analysis producing the fit considers a perfectly stable phase reference for the setup (lock-in plus generator). Only a base-line is subtracted, with a resistive (on $V'$) and a reactive (on $V''$) component. Both components are taken to be linear with respect to the frequency sweep, that is either the background is a linear function of $f$, or its magnitude changes weakly with time. For narrow sweeps or small temperature drifts, this procedure is perfectly justified.
The fits are excellent. The small discrepancies in Fig. \ref{3lines} are mostly due to the finite-time needed to sweep the line (especially at the bifurcation point) or drifts in the circuit parameters. Limitations of the model are discussed in part \ref{last}.

\section{THEORETICAL DESCRIPTIONS}
\label{theor}

In this section we present the theoretical approaches used in order to quantitatively describe our results. The aim is to use {\em exclusively} analytical expressions, without any numerical simulations. The simplifications we apply are thoroughly discussed.\\
We shall start by introducing the required formulas specialized to our structure, and pursue with a refined application of the Rayleigh method to the non-linear resonance case. A discussion of the friction signature is presented.

\subsection{Simple static distortion}
We first consider the simplest case of a static linear distortion, for a straight ($\alpha=0$) and ideally thin structure. 
We take into account the presence of (homogeneous) stress in the Silicon, along the feet and the paddle, which can arise from the metal deposition$^($\footnote{In one very thin sample (less than 5 $\mu$m) covered with NbTi, the authors even noticed under the SEM a strong static distortion of the paddle of the structure. The influence of the metal layer is discussed in detail in the oncoming experimental sections.}$^)$. It is modeled with two static forces $S$, one along $y$ (paddle) and one along $z$ (feet), which sign depend on the stress type (tension or compression). \\
The aim is to demonstrate that the {\em lowest mode} of the structure is simply the back-and forth oscillation of the feet clamped together by the paddle. This mode is the one we presented in part \ref{experiment}, and study quantitatively in part \ref{datanalys}. 

\subsubsection{Paddle distortion}
\label{padstatic}
In order to calculate the shape $X(y)$ of the paddle distortion, we assume that the clamping of the paddle with respect to the feet is ideal ($X=0$, $dX/dy=0$ at $y=0,l$). The force is taken to be homogeneously distributed along the paddle ($d F / dl =I B  = \,$Cst). The results are\cite{timoRDM}:
\begin{eqnarray*}
X(y) & = & \!\! + l \left(\frac{F l^2}{E_y I_y}\right) \frac{1}{u^4}\left[ \frac{+u/2}{\tanh(u/2)} \left(\frac{\cosh u(1/2-y/l)}{\cosh(u/2)}-1\right) + \frac{u^2}{2} \frac{y (l-y) }{l^2}\right]  , \\
 & or &\\
X(y) & = & - l \left(\frac{F l^2}{E_y I_y}\right) \frac{1}{u^4}\left[ \frac{-u/2}{\tan(u/2)} \left(\frac{\cos u(1/2-y/l)}{\cos(u/2)}-1\right) + \frac{u^2}{2} \frac{y (l-y) }{l^2}\right]  \\
\end{eqnarray*}
where $u=\sqrt{\frac{\left|S\right| l^2}{E_y I_y}}$ and the first expression stands for tension
($S>0$ with our notations) and the second for compression ($S<0$). $F$ is the total force ($I l B$), $E_y$ stands for the Young's modulus along the $y$ axis, and $I_y$ the 
corresponding moment of inertia for flexion. With our notations, we get $I_y=\frac{1}{12} a \, e^3$. 

\vspace{0.15in}
Assuming that the stress $S$ is small with respect to $E_y I_y /l^2 $ ($u \ll 1$), we write:
\begin{eqnarray}
X(y)  & = & X_m \Biggl[ \left( +16 (y/l)^2 - 32 (y/l)^3 +16 (y/l)^4 \right) \nonumber \\
 & \pm &\!\! u^2 \left( \frac{ +2 (y/l)^2 -12 (y/l)^3 + 26 (y/l)^4 -24 (y/l)^5 +8 (y/l)^6 }{15} \right) \Biggr] \label{eqpaddlestatic} \\
 \nonumber
\end{eqnarray}
where $X_m = l \left(\frac{F l^2}{E_y I_y}\right) \left[  \frac{1}{384}-\frac{\pm u^2}{15360} \right] $ is the maximal deflection at $y=l/2$. The elongation of the paddle due to the force $F$ has a second order signature on the axial stress $S$, which is neglected here. The $\pm$ stands for the sign of $S$; the first order term in stress is a quadratic $u^2$.  \\
The bending moment $M_z$ at each end of the paddle is then:
\begin{equation}
\left| M_z \right| = \left| F \right| l \left( \frac{1}{12} - \frac{\pm u^2}{720} \right).
\label{bend}
\end{equation}
The force exerted at the ends has an axial component $\pm S$ along $y$ (up to a second order in $F$), and a shear component of magnitude $ \left|F\right|/2$ along $x$. \\
The stress tensor $\underline{\sigma}$ in the Silicon has only three nonzero components:
\begin{eqnarray}
\sigma_{yy}(x,y) & = & \frac{S}{e \, a} + \frac{ 12 \, F}{e \, a} \frac{l}{e} \left[ (x/e)- \frac{1}{2} \right] 
\Biggl[ \left(\frac{1}{12} - \frac{(y/l)}{2}+ \frac{(y/l)^2}{2}\right) \nonumber\\
 &\pm & u^2 \left( -\frac{1}{720}+ \frac{(y/l)^2}{24}-\frac{(y/l)^3}{12}+\frac{(y/l)^4}{24} \right) \Biggr] ,  \label{syypad} 
\end{eqnarray}
\begin{eqnarray}
\sigma_{yx,xy}(x,y) & = & \frac{ 3 \, F}{e \, a} \Biggl[ \left( -(x/e)+ (x/e)^2 \right) \Bigl( 1 - 2 (y/l) \Bigr) \nonumber\\
& \pm & u^2 (x/e) (y/l) \, \Bigl( 1 -  (x/e) \Bigr)  
\left( \frac{1-3(y/l)+2(y/l)^2 }{6}\right) \Biggr] \label{syxpad}
\end{eqnarray}
where the double subscript $yx,xy$ recalls the tensor symmetry ($x$ running from 0 to $e$).

\vspace{0.15in}
If we assume that there is no slippage at the metal/Silicon interface, one can easily take into account the finite Young's modulus of the metal layer. The actual structure is then equivalent to an all-Silicon "{\it T}-shaped" beam, where the width $\tilde{a}$ of the top part (of thickness $e_M$) is:
\begin{displaymath}
\tilde{a} = a \frac{E_M}{E_y}  \\
\end{displaymath}
where $E_M$ is the Young's modulus of the metal layer. By doing so, one has to replace in the above equations the moment of inertia by:
\begin{equation}
I_y \longrightarrow \frac{1}{12} \frac{a^2 e^4 + \tilde{a}^2 e_M^4 + 4 \, a e \, \tilde{a} e_M (e^2 + e_M^2 + \frac{3}{2} e \, e_M)}{a e + \tilde{a} e_M} .\label{eqpaddleIz}
\end{equation}
In our case (using mostly Aluminum), typically $E_M/E_y \approx 0.4$ and $e_M/e < 0.15 $, thus the correction is small, but measurable.

\subsubsection{Feet distortion}
\label{feetstatic}
We proceed in the same way for the feet. The shape $X(z)$ of each foot's distortion is calculated by assuming a perfect clamping of the foot at its basis ($X=0$, $dX/dz=0$ at $z=0$). The force $F$ is taken to be applied at the end of the foot, where the paddle is anchored. The results are\cite{timoRDM}:
\begin{eqnarray*}
\lefteqn{X(z)  = } \\
& + h \left( \frac{F h^2}{E_z I_z} \right) &   \frac{1}{u^3 (u-1 + \cosh u)} \biggl[ \, +  u (1-u) (1-z/h)  \\
& + &  \cosh (u)\, (u \, z/h- \tanh(u) )  +  (1-u)(\sinh (u \,z/h) -u \cosh(u\, z/h)) \\ 
& + &  \sinh \, u (1-z/h) \, \biggr]   , \\
 & or &\\
 & & \\
\lefteqn{X(z)  = }  \\
& - h \left( \frac{F h^2}{E_z I_z} \right) &   \frac{1}{u^3 (u-1 + \cos u)} \biggl[ \, +  u (1-u) (1-z/h)  \\
& + &  \cos (u) \,(u \, z/h- \tan(u) )  +  (1-u)(\sin (u \,z/h) -u \cos(u\, z/h)) \\ 
& + &  \sin \, u (1-z/h) \, \biggr]   \\
\end{eqnarray*}
where $u=\sqrt{\frac{\left|S\right| h^2}{E_z I_z}}$ and the first expression 
stands for tension ($S>0$) and the second for compression ($S<0$). From the previous section, the notations are straightforward, and we have $I_z=\frac{1}{12} e' \, e^3$. 

\vspace{0.15in}
Assuming that the stress $S$ is small with respect to $E_z I_z /h^2 $ ($u \ll 1$), we can again develop the expressions and write:
\begin{eqnarray}
X(z)  & = & X_m \Biggl[ \left(  \frac{3}{2} (z/h)^2 - \frac{1}{2} (z/h)^3 \right) \nonumber \\
 & \pm & u \left( \frac{ (z/h)^2 - (z/h)^3 }{4} \right)  \Biggr] \label{eqfootstatic} \\
 \nonumber
\end{eqnarray}
where $X_m = h \left(\frac{F h^2}{E_z I_z}\right) \left[  \frac{1}{3}-\frac{\pm u}{6} \right] $ is the maximal deflection at $z=h$. The $\pm$ stands again for the sign of $S$. Note the {\em linear} dependence to $u$.
The bending moment $M_y$ at the clamping end of the foot is then:
\begin{equation}
 M_y  = F h \left( 1- \frac{\pm u}{3} \right)
\end{equation}
while the shear force exerted (along $x$) at the same end is of course $-F$, and the axial force (along $z$) is by definition $-S$. \\
The stress tensor $\underline{\sigma}$ in the Silicon has only three nonzero components:
\begin{eqnarray}
\sigma_{zz}(x,z) & = & \frac{S}{e \, e'} + \frac{ 12 \, F}{e \, e'} \frac{h}{e} \left[ (x/e)- \frac{1}{2} \right] \left[1- (z/h) -\frac{\pm u}{3}\right] , \label{szzfoot} \\
\sigma_{zx,xz}(x,z) & = & \frac{6 \, F}{e \, e'} \left[  -(x/e) +(x/e)^2  \right] \label{szxfoot}
\end{eqnarray}
where the double subscript $zx,xz$ recalls the tensor symmetry (and $x$ runs from 0 to $e$).

\vspace{0.15in}
Assuming that there is no slippage at the metal/Silicon interface, we treat the finite Young's modulus of the metal layer $E_{M}$ as in the previous paragraph. Introducing $\tilde{e'}=e'\, E_M/E_z$, the moment of inertia becomes:
\begin{equation}
I_z \longrightarrow \frac{1}{12} \frac{e'^2 e^4 + \tilde{e'}\,\!^2 e_M^4 + 4 \, e' e \, \tilde{e'} e_M (e^2 + e_M^2 + \frac{3}{2} e \, e_M)}{e' e + \tilde{e'} e_M} \label{eqfootIz}
\end{equation}
which is again a small correction to the standard $I_z$.

\subsubsection{Structure behavior}
\label{rayleigh}

We shall see in section \ref{forced} that for the first mode of vibration of our beams (feet or paddle), the peak deflection $X_{max}$ of the maximal displacement $ X_m (t) $ obtained at resonance by an {\it ac} force of peak value $F_{ac}$, is equal$^($\footnote{Within the approximation identifying the static spring constant (and the $k_v$ defined in this section) and the normal spring constant $k_n$ of the first mode, which amounts to $\pm 3\,$\% discrepancies maximum.}$^)$ to the static deflection $X_m$ obtained with a force $F_{dc}$:
\begin{displaymath}
F_{dc}=Q F_{ac}
\end{displaymath}
where $Q$ is the quality factor of the mode resonance. Besides, the static shapes Eq. (\ref{eqpaddlestatic}) and (\ref{eqfootstatic}) are fairly good approximations of the dynamical shapes. \\
Indeed, one can calculate approximative dynamical parameters of the first paddle and feet modes from these shapes. This energetic method, called the Rayleigh method\cite{timoVIBR}, consists in integrating the elastic energy and kinetic energy along the beam, with the time dependence 
incorporated in $X_m \rightarrow X_m(t)$. For instance with the foot distortion:
\begin{eqnarray*}
\frac{1}{2} k_{v} \, X_m(t)  ^2 & = & \frac{1}{2} \int_{0}^{h}  \left( \frac{d^2 X(z,t)}{d z^2} \right)^2  (E_z I_z) \, dz ,\\
\frac{1}{2} m_{v} \left( \frac{dX_m(t)}{dt} \right)^2 & = & \frac{1}{2} \int_{0}^{h}  \left( \frac{d X(z,t)}{d t} \right)^2 
( \rho e e' ) \, dz
\end{eqnarray*}
which define respectively the spring constant $k_v$ and the vibrating mass $m_v$ associated to each foot (for the first mode). $\rho$ is the mass density of Silicon. The same $k_v$ can be derived by writing the constitutive equation of the elastic force at the end of the foot: $F_{dc} = k_v \, X_m$. These equations are easily transposed to the paddle case.

\vspace{0.15in}
In order to take into account the metallic layer, one has to change $I_z$ according to the previous section, and also to replace $\rho e \rightarrow \rho e + \rho_M e_M$ in the above equations ($\rho_M$ being the metal density).

\vspace{0.15in}
We shall see that this knowledge is enough to produce a good description of 
the {\em global} shape of the dynamically distorted structure.  \\
Consider first the paddle's clamping hypothesis. Due to the stress $S$ along the paddle, each foot will tend to have a distortion along the $y$ axis (adapting Eq. (\ref{eqfootstatic})):
\begin{displaymath}
\left|Y_m \right| =  h \left(\frac{\left| S \right| h^2}{E_z I'_z}\right) \left[  \frac{1}{3}-\frac{\pm u}{6} \right]
\end{displaymath}
where $I'_z=\frac{1}{12} e \, e'^3$ (note the inversion between $e'$ and $e$ from $I_z$; the $u$ here is the one generated by the stress {\em along the foot}). Depending on the sign of $S$, the feet either bend inwards or outwards. \\
The bending moment Eq. (\ref{bend}) at each end of the paddle tends to twist the feet by an angle:
\begin{displaymath}
\left| \theta \right| = \left| F_{dc} \right| \, l \left( \frac{1}{12} - \frac{\pm u^2}{720} \right) \frac{h}{G_z J_z}
\end{displaymath}
where $G_z$ is the Silicon shear modulus (along $z$) and $J_z=\frac{1}{12} (e' e^3 + e e'^3)$ the corresponding moment of inertia ($u$ is here the one associated to the stress {\em along the paddle}). Obviously, the feet's twist is always oriented toward the bending force $F_{dc}$.\\
Using the typical geometrical dimensions of our devices (say, sample E6) and the Silicon $E_z$ and $G_z$ values\cite{silicon} (160$\,$GPa and 80$\,$GPa respectively), we get for the maximum solicitation$^($\footnote{These estimates neglect non-linearities; the same conclusions are reached with the non-linear expressions of section \ref{nonlin}.}$^)$ we consider ($S \approx \,$60 $\,\mu$N and $F_{dc} \approx\,$0.4$\,$mN, part \ref{datanalys}):
\begin{eqnarray*}
\left|Y_m \right|     & < & 1 \%    \,\, h    ,  \\
\left| \theta \right| & < & 2^\circ  .
\end{eqnarray*}
These are negligible (even when the resonance becomes highly non-linear), and we can safely state that the clamping hypothesis of the paddle is justified. \\
Consider now the distortion of the paddle {\em itself} under the solicitation $F_{dc}$. The maximum deflection in the middle is (applying Eq. (\ref{eqpaddlestatic})):
\begin{displaymath}
X_m = l \left(\frac{F_{dc} \, l^2}{E_y I_y}\right) \left[  \frac{1}{384}-\frac{\pm u^2}{15360} \right] .
\end{displaymath}
Using the maximum values of part \ref{datanalys} we get:
\begin{displaymath}
\left| X_m \right| < 1 \% \,\,l 
\end{displaymath}
which, again, is negligible, even at the largest distortions. We can thus state that, as far as the feet's flexion is concerned, the paddle is always perfectly rigid.

\vspace{0.15in}
These considerations allow to {\em identify} simply the two first modes of the whole structure. One is the flexion of the feet together with the rigid paddle, and one is the oscillation of the paddle at the end of rigid feet (which, according to the above value of $X_m$, is very poorly excitable).

\vspace{0.15in}
Using the Rayleigh method we can approximate these two modes. 
The vibrating mass and spring constant of the feet give an angular frequency $\omega_0 = \sqrt{2 k_v/(2 m_v + m_{paddle})}$ for their first mode (the feet are loaded by the paddle mass $m_{paddle}$). 
For the paddle, the same procedure gives an angular frequency $\omega_0 = \sqrt{k_v/m_v}$ for the first clamped mode. \\
In Tab. \ref{tabRay} we summarize the expressions obtained within this approximation. Comparing with section \ref{harmonic} the prefactors $2\sqrt{35/11}$ and $6\sqrt{14}$ give fairly good estimates of the mode positions. However, the influence of the stress term $u$ {\em fails to be described} correctly. Thus, an accurate modeling of the linear response of the system requires the exact time-dependent solution given in the next section.

\begin{table}[h!]
\begin{center}
\hspace*{-0.10 cm}
\begin{tabular}{|c|c|c|}    \hline
Property &  For (loaded) feet & For (clamped) paddle     \\    \hline    \hline
$k_v$    & 3  $E_z I_z / h^3$ & $\frac{1024}{5} \left(1 + \frac{\pm u^2}{420} \right)$ $E_y I_y / l^3$ \\   \hline
$m_v$    & $\frac{33}{140} \left(1 - \frac{\pm 8 u}{99} \right)$  $\rho e e' h$ & $\frac{128}{315} \left(1 + \frac{\pm u^2}{660} \right)$  $\rho e a l$    \\   \hline    \hline
$\omega_0$ & $2\sqrt{\frac{35}{11}}\sqrt{\frac{1}{1+ 70/33 \, \eta}} \left( 1 + \frac{\pm 4 u }{99 + 210 \, \eta} \right)$ $\sqrt{\frac{E_z I_z/ h^3}{\rho e e' h}}$ & $6\sqrt{14}\left(1 + \frac{\pm u^2}{2310} \right) $ $\sqrt{\frac{E_y I_y/ l^3}{\rho e a l}}$\\   \hline
\end{tabular}
\caption{\label{tabRay}Rayleigh's method parameters (see text). $\pm$ denotes the stress sign, source of the term $u$. For the loaded feet, we introduced the reduced parameter $\eta = (a l)/(e' h)$ (ratio of paddle mass to foot mass). 
The metal contribution has been omitted for clarity.}
\end{center}
\end{table}

\subsection{Simple harmonic treatment}
\label{harmo}

We calculate in this section the normal modes arising in the linear regime, for a straight structure ($\alpha=0$). 
As in the static case, we take into account the presence of (homogeneous) stress  in the Silicon through $S$, along the feet or the paddle (with the same sign convention). We solve then the forced oscillations of the feet around the first mode. 

\subsubsection{Normal modes}
\label{harmonic}
Consider the distortion of one foot $X(z,t)$. The equation one has to solve is\cite{timoVIBR}:
\begin{displaymath}
E_z I_z \frac{\partial^4 X}{\partial z^4} = \pm S \frac{\partial^2 X}{\partial z^2} - \rho \, e e' \frac{\partial^2 X}{\partial t^2}
\end{displaymath}
where we used the same notations as in the above sections. We seek solutions (the normal modes) of the form:
\begin{displaymath}
X(z,t) = X(z) \cos (\omega_0 t + \varphi)
\end{displaymath}
where $\varphi$ is an arbitrary constant (depending on the choice of the time reference) and $\omega_0$ has to be defined.\\
We introduce again the parameter $u= \sqrt{\left|S \right| h^2 /(E_z I_z)}$, and assume in the following that $u \ll 1$. One can show that the clamped condition at the foot's end ($X=0$, $dX/dz=0$ at $z=0$) together with a "half-free" condition on the other side (zero bending moment, $d\,^2 X/dz^2=0$ at $z=h$) brings the solution:
\begin{eqnarray}
\lefteqn{X(z) = \frac{X_m}{c_1(\lambda_{-},\lambda_{+})} \biggl[ 
\Bigl( \sin (\lambda_{-}\frac{z}{h})-\frac{\lambda_{-}}{\lambda_{+}} \sinh (\lambda_{+}\frac{z}{h} )  \Bigr) } \nonumber \\
&&+ c_2(\lambda_{-},\lambda_{+}) 
\Bigl(\cos (\lambda_{-}\frac{z}{h})-\cosh (\lambda_{+}\frac{z}{h} ) \Bigr) \biggr]
\label{normalmode}
\end{eqnarray}
with:
\begin{eqnarray*}
c_1(\lambda_{-},\lambda_{+}) & = & \frac{\lambda_{-}^2+\lambda_{+}^2}{\lambda_{+}} \left(
\frac{\lambda_{+} \cosh(\lambda_{+}) \sin (\lambda_{-}) -\lambda_{-} \cos(\lambda_{-})\sinh(\lambda_{+}) }
{\lambda_{-}^2 \cos(\lambda_{-}) + \lambda_{+}^2 \cosh (\lambda_{+})} \right) ,\\
c_2(\lambda_{-},\lambda_{+})& = & - \lambda_{-} \left( \frac{\lambda_{-} \sin(\lambda_{-})+\lambda_{+} \sinh(\lambda_{+})}
{\lambda_{-}^2 \cos(\lambda_{-}) + \lambda_{+}^2 \cosh (\lambda_{+})} \right)
\end{eqnarray*}
and:
\begin{eqnarray*}
\lambda_{-} & = & \lambda \left( 1 - \frac{1}{4} \, \frac{\pm u^2}{\lambda^2} \right)  ,\\
\lambda_{+} & = & \lambda \left( 1 + \frac{1}{4} \, \frac{\pm u^2}{\lambda^2} \right)  ,\\
\lambda & = & \left( \frac{  \rho e  e'\, \omega_0^2 \, h^4}{E_z I_z} \right)^{1/4}
\end{eqnarray*}
where the $\pm$ sign stands for tension ($S>0$) or compression ($S<0$). Note that we take into account the finite thickness of the metal layer with the same change of $\rho e$ and $I_z$ than in the previous section. In the above equations, $\lambda$ has to be determined by means of the last boundary condition:
\begin{displaymath}
E_z I_z \frac{d^3 X}{d z^3}(h) = - \frac{m_{paddle}}{2}  \, \omega_0^2 \, X(h)
\end{displaymath}
which states that the shear force at the end $z=h$ is due to the inertia of the foot's load (which is, due to symmetry, half of the paddle's mass inertia$^($\footnote{A slight asymmetry between the feet {\em only} shifts the paddle mass-per-foot one has to consider in this calculation.}$^)$ per foot). The solution can be written in the form:
\begin{equation}
\lambda = \lambda_i(\eta) \left( 1 - \frac{\pm u^2}{2} \, \phi_i(\eta) \right) \label{mode}
\end{equation}
where $\lambda_i(\eta)$ is the solution without stress ($u=0$) and $\phi_i(\eta)$ is the sensitivity to stress (for mode $i$). Both depend on the ratio $\eta= (a l)/(e' h)$ of the paddle mass to the foot mass. Indeed, the mode position is highly sensitive\cite{masspaper} to the end load $m_{paddle}/2$.\\
These functions can be determined numerically for each mode; in particular for the first one $i=0$ good asymptotic expressions are:
\begin{eqnarray*}
\lambda_0(\eta) & = & \left[ \frac{(\lambda_{00})^4/4 + 3 \, \eta/2 }{1 + \eta/2} \, \frac{1}{1/4 + \eta/2}\right]^{1/4} ,\\
\phi_0(\eta) & = & 0.034712 + 0.21210 \,\, \eta/2 ,\\
\lambda_{00} & = & 1.87510
\end{eqnarray*}
with $\lambda_{00}$ the well-known value assigned to the first mode of the ideal cantilever. The first expression fits within 0.1 \% and the second within 0.5 \% at worst. 
This approach considered a bulk stress source $S$ {\em without stipulating its origin}. We refer to the discussion of part \ref{datanalys} dealing with the (controversial) issue of surface stresses\cite{surfacestress}, and our related quantitative analysis.\\
The corresponding solutions can be produced for the paddle, with clamped conditions on both sides (and of course no load). One finds then 
$\lambda_{00}=4.7300$ which is the well-known position of the first mode of the ideally clamped beam. \\
With the experimental parameters of part \ref{datanalys}, one can then easily verify that the feet mode, which is the one of interest to us, and the (weakly excitable) paddle mode are very well separated.

\vspace{0.15in}
From the above expressions, one can interestingly notice that as the ratio $\eta$ increases, the dynamical shape becomes closer and closer to the static one we obtained in section \ref{feetstatic}; as far as the {\em shape} of the distortion is concerned, the Rayleigh method appears to be rather accurate. Indeed, comparing these first mode positions to the ones obtained in the previous section by the Rayleigh method (Tab. \ref{tabRay}), we see that the disagreement is always smaller than 1.5$\,$\% (and falls to zero as $\eta \rightarrow +\infty$). However, note the differences in the incorporation of stress (through $u$) in the solution.

\subsubsection{Forced oscillation}
\label{forced}

In order to compute the response of the system to a harmonic force $F(t)=F_{ac} \cos (\omega t)$ of (arbitrary) angular frequency $\omega$, we apply the method of the virtual work\cite{timoVIBR}. \\
Take for the (general) distortion of one foot the expression:
\begin{displaymath}
X(z,t)=\sum_{i=0}^{+\infty} X_{m}^{(i)}(t) \, X_i (z)
\end{displaymath}
and for the virtual displacement:
\begin{displaymath}
\delta  (z,t)=\Delta (t) \, X_j (z)
\end{displaymath}
where the $X_l (z)$ functions are the (normalized) normal modes$^($\footnote{That is Eq. (\ref{normalmode}) from the previous section with $X_m=1$, and $\lambda$ equal to the mode value $\lambda_l$, Eq. (\ref{mode}).}$^)$ of the foot and $X_{m}^{(i)}(t)$, $\Delta (t)$ time-dependent (maximal) deflections. The $X_{m}^{(i)}(t)$ have to be defined, and the $\Delta (t)$ shall disappear at the end of the calculation.\\
Using these expressions we write the terms appearing in the power balance. We get, integrating along $z$, for one foot:
\begin{eqnarray*}
\delta W_I^j & = & - (\rho e e' ) \, \Delta (t) \, \sum_{i=0}^{\infty} \ddot{X}_{m}^{(i)}(t) \,\,  h \, \gamma_{i,j}    , \\
\delta W_E^j & = & - (E_z I_z) \, \Delta (t) \, \sum_{i=0}^{\infty} X_{m}^{(i)}(t)   \,\,   \tilde{\gamma}_{i,j} /h^3
\end{eqnarray*}
where the first is the inertia term, and the second the elastic term (dots are time-derivatives). 
For the end load acting on one foot we get:
\begin{eqnarray*}
\delta W_{I,m_{paddle}}^j & = & - \frac{1}{2} m_{paddle} \, \Delta (t) \, \sum_{i=0}^{\infty}   \ddot{X}_{m}^{(i)}(t) \,\, X_i (z=h) X_j (z=h)   , \\
\delta W_{E,m_{paddle}}^j & = & + E_z I_z \, \Delta (t) \, \sum_{i=0}^{\infty} X_{m}^{(i)}(t) \,\,  X ''' _i(z=h) X_j (z=h)
\end{eqnarray*}
where the first is the inertia term (of the half-paddle), and the second the restoring force. One has of course to add the work due to the external force:
\begin{displaymath}
\delta W_F^j  =  + F(t) \, \Delta (t) \, X_j (z=h)  .
\end{displaymath}
In the above expressions, one has to make use of the definition
$X_i (z=h)=X_j (z=h)=1$. In order to keep the equations simple, we will neglect in the following the dependence to stress$^($\footnote{Indeed, the stored stress $S$ couples the normal modes (see discussion of part \ref{last}). Moreover, extracting $S$ from the measurements is a non-trivial issue (sections \ref{dissipe} and part \ref{datanalys}).}$^)$ ($u$=0). We introduced the quantities:
\begin{eqnarray*}
\tilde{\gamma}_{i,j}  =  h^3 \int_{0}^{h} \frac{d^4 X_i}{d z^4}(z) X_j (z) dz  =  + \lambda_i^4 \, \, \gamma_{i,j}  , \\
X ''' _i(z=h)  =  \frac{d^3 X_i}{d z^3}(z=h)  =  \frac{-\eta/2 \, \,\lambda_i^4  }{h^3}
\end{eqnarray*}
together with:
\begin{displaymath}
\gamma_{i,j}  = \frac{1}{h} \int_{0}^{h} X_i (z) X_j (z) dz  
\end{displaymath}
which verifies:
\begin{eqnarray*}
\gamma_{i,j} & = & \gamma_j   \,\,\ { when  } \,\,\ i=j ,\\
&  = & - \eta/2   \,\,\ { when } \,\,\ i\neq j .
\end{eqnarray*}
Adding up all the terms of the power balance and using the equations above, one finally gets, for the whole structure:
\begin{equation}
2 \left[ \rho e e' h \, \gamma_{j} + \frac{\rho e a l}{2} \right]  \ddot{X}_m^{(j)}(t) + 2 \frac{E_z I_z}{h^3}  \, \lambda_j^4 \, \left[  \gamma_{j} +  \frac{\eta}{2} \right] X_m^{(j)}(t)= F(t).
\label{equa}
\end{equation}
We recognize the simple one-dimensional forced oscillator expression with $k_n^{(j)}=  (\frac{E_z I_z}{h^3})  \, \lambda_j^4 \, \left[  \gamma_{j} +  \frac{\eta}{2} \right]$ (normal spring) and $m_n^{(j)} =  (\rho e e' h) \left[ \gamma_{j} + \frac{\eta}{2} \right] $ (normal mass) associated to each foot. \\
A good asymptotic expression for the first mode $\gamma_{j}=\gamma_{0}(\eta)$ is (again neglecting the stress term $u$):
\begin{displaymath}
\gamma_0 (\eta) =  \frac{\frac{1}{4} +  \frac{33}{140} \,\, 4.418 \, \eta/2}{1 + 4.418 \, \eta/2}
\end{displaymath}
which fits the numerical result within 0.1$\,$\%.

\vspace{0.15in}
In section \ref{dissipe} we will model friction through a dissipation term proportional to speed, namely $-2 \Lambda_1 \, \dot{X}_m^{(j)} (t)$, and a reactive term proportional to acceleration, $-2 \Lambda_2 \, \ddot{X}_m^{(j)} (t)$.
Injecting in Eq. (\ref{equa}) and solving for the harmonic $F(t)$, the expression of $X_m^{(j)} (t)$ becomes:
\begin{eqnarray}
\!\!\!\!\!\!\!\!\!\!\!\!\! X_m^{(j)}(t) & = &  X\!P_m^{(j)} \cos (\omega t) + X\!Q_m^{(j)} \sin (\omega t) \nonumber \\
& with & \nonumber \\
\!\!\!\!\!\!\!\!\!\!\!\!\! X\!P_m^{(j)}    & = &  \frac{F_{ac}}{2k_n^{(j)}} \frac{1-(\omega/\omega^{(j)}_{0})^2- 2 \delta \omega \, \omega /(\omega^{(j)}_0 )^2}{\left[1-(\omega/\omega^{(j)}_0)^2- 2\delta \omega \,\omega /(\omega^{(j)}_0)^2\right]^2+ \left[\Delta \omega \, \omega/(\omega^{(j)}_0)^2\right]^2} , \label{phase}\\
\!\!\!\!\!\!\!\!\!\!\!\!\!  X\!Q_m^{(j)}    & = &  \frac{F_{ac}}{2k_n^{(j)}} \frac{\Delta \omega \, \omega/(\omega^{(j)}_0)^2}{\left[1-(\omega/\omega^{(j)}_0)^2- 2\delta \omega \,\omega /(\omega^{(j)}_0)^2\right]^2+ \left[\Delta \omega \, \omega/(\omega^{(j)}_0)^2\right]^2} \label{quad}
\end{eqnarray}
where $X\!P_m^{(j)}$ is the (peak) in-phase motion, and $X\!Q_m^{(j)}$ the (peak) out-of-phase motion (with $\omega^{(j)}_{0}=\sqrt{k_n^{(j)}/m_n^{(j)}}$ the angular frequency of mode $j$). We introduced $\delta \omega= \omega \Lambda_2 /2m_n^{(j)}$ and $\Delta \omega=2 \Lambda_1 / 2m_n^{(j)}$; note that in the most general case $\Lambda_1$ and $\Lambda_2$ are functions of $\omega$ (section \ref{dissipe}). 
In the Fourier language, we write $\tilde{X}_m^{(j)}(\omega)=X\!P_m^{(j)}- i \, X\!Q_m^{(j)}$ (with $i= \sqrt{-1}$).

\vspace{0.15in}
Equations (\ref{phase}) and (\ref{quad}) are characteristic of a resonance. Considering the under-damped regime ($\delta \omega \ll \omega_0^{(j)}$ and $\Delta \omega \ll \omega_0^{(j)}$ together with $\Lambda_1$ and $\Lambda_2$ constant), these relations reduce to the well-known {\em Lorentz} shape:
\begin{eqnarray*}
X\!P_m^{(j)} & = & \frac{F_{ac}}{2k_n^{(j)}} \, \frac{+2 \omega_0^{(j)} (\omega_{res}-\omega)}{4 (\omega_{res}-\omega)^2+(\Delta \omega)^2}   , \\
X\!Q_m^{(j)} & = & \frac{F_{ac}}{2k_n^{(j)}} \,\frac{+ \Delta \omega \, \omega_0^{(j)}}{4 (\omega_{res}-\omega)^2+(\Delta \omega)^2}
\end{eqnarray*}
with $\omega_{res}=\omega_0^{(j)}- \delta \omega $. At resonance $\omega=\omega_{res}$, we have $X\!P_m^{(j)}=0$ and $X\!Q_m^{(j)}=X_{max}= \frac{F_{ac}}{2k_n^{(j)}} \, Q^{(j)}$ with $Q^{(j)}=\omega_0^{(j)}/\Delta \omega$: the displacement is out-of-phase, with a peak value amplified by the quality factor $Q^{(j)}$. The full-width-at-half-height of the function $X\!Q_m^{(j)}$ is $\Delta \omega$. \\
This is of course fairly general. Equivalent equations can thus be obtained for the paddle vibrations. However in the following, we will be interested only in the first $j=0$ feet mode, and shall thus drop the superscript $(j)$. Note that the normal spring constant $k_n$ in the first mode is practically identical to the static one (or to $k_v$ obtained in the Rayleigh calculation, see section \ref{rayleigh}).

\vspace{0.15in}
The signal we measure is $V=  B l \, \vec{z}.\vec{\mathsf{z_m}} \,\,d X_m/\!dt$. In the linear regime, the angle between $\vec{z}$ and the direction $\vec{\mathsf{z_m}}$ along the distorded feet, at their end point, is negligible and 
$\vec{z}.\vec{\mathsf{z_m}} \approx 1$. Taking the time derivative of $X_m (t)$ one easily gets the Lorentzian expression used to fit the Fourier components $\tilde{V}(\omega)$ in the (steady) linear regime. \\
The non-linear regime requires further modeling which is described in the next section.

\subsection{Non-linear Rayleigh method}
\label{nonlin}

As the driving force is increased, one eventually enters into the {\em non-linear regime}. In Eq. (\ref{equa}), the normal spring constant and normal mass become functions of $X_m(t)$. This is, in general, also true for the friction terms ($\Lambda_1$ and $\Lambda_2$, section \ref{dissipe}) and the detected signal (through the $\vec{z}.\vec{\mathsf{z_m}}$ term). \\
Two types of non-linearities can be distinguished:
\begin{itemize}
\item Non-linearities due to {\em the materials} which are used beyond their elastic limit, or displaying non-linear friction mechanisms.
\item Non-linearities originating from {\em the geometry} of the structures.
\end{itemize}
The friction's non-linearities shall be addressed in section \ref{dissipe} and discussed on the basis of our experimental results in part \ref{datanalys}. The elastic limit of Silicon can be compared to the maximum stress we compute for the structure (using $S \approx \,$60$\,\mu$N and $F_{dc} \approx\,$0.4$\,$mN, part \ref{datanalys}):
\begin{center}
Paddle: \\
$\sigma_{yy} \,\,\,\,\, < 1$ GPa       $\,\,\,\,\,$ Eq. (\ref{syypad})  ,\\
$\sigma_{yx,xy} < 1$ MPa               $\,\,\,\,\,$ Eq. (\ref{syxpad})  ,
\end{center}
\begin{center}
Foot: \\
$\sigma_{zz}  \,\,\,\,\, < 1$ GPa      $\,\,\,\,\,$ Eq. (\ref{szzfoot}) ,\\
$\sigma_{zx,xz}  < 2$ MPa       $\,\,\,\,\,$ Eq. (\ref{szxfoot}) .\\
\end{center}
The (self-consistent) use of the non-linear results of this section brings the same conclusions. \\
Monocrystalline Silicon is brittle, and rupture occurs in the linear regime; there is no plastic yield limit. Values for the failure stress differ from one sample to the other, but typical values\cite{silicon} quoted in the literature are {\em above} $1\,$GPa.
The Silicon structure is thus elastic in the whole studied excitation range. \\
However, the metallic layer on top of the sample will {\em not} remain linear\cite{paperthinAl} at strong drives (since the metal yield strength may be as low as 300$\,$MPa). Because the metal contribution is very small, it shall not contribute dominantly to the non-linear behavior of the structure. However, plasticity of the metallic layer will give rise to {\em permanent frequency shifts} when the force is decreased again down to the linear regime (by changing the permanent stress term $u$, see part \ref{last}).

\vspace{0.15in}
We are consequently left with the geometric non-linear effects. In order to calculate estimates of the non-linear parameters, we will extend the Rayleigh method to the non-linear case. To our knowledge, this is the simplest way to produce an {\em analytical} description. Since in the linear regime, the shape of the distortion was fairly well reproduced by this method, a semi-quantitative agreement with experiments is expected.  

\vspace{0.15in}
We start by solving the foot's non-linear static case. We do not take into account additional stress $S$ stored in the structure ($u=0$) and neglect in Hooke's law the Poisson's ratios (which would be of order $0.2-0.3$ for Silicon\cite{silicon}). We consider a finite thickness with a gradient $\alpha$, and neglect deviations in the thickness expression $e(z)$ to linearity when a force is applied.
We checked that these simplifications have no impact on the discussion presented here. \\
The sole source of geometrical non-linearity originates in the expression of the radius of curvature $r(z)$ (of the top of the structure $X(z)$, metal-covered in Fig. \ref{fig1}):
\begin{displaymath}
\frac{1}{r(z)} = \frac{X''(z)}{[ \, 1+X'(z)^2 \, ]^{3/2}}
\end{displaymath}
where the primes denote derivation with respect to $z$. We restrict the calculation to third order terms in the normalized applied force $F_{norm}=\frac{F h^2}{E_z I_z(0)}$ (with $I_z(0)=\frac{1}{12} e' e(0)^3$ the moment of inertia at the clamping end), which brings:
\begin{displaymath}
\frac{1}{r(z)} = X''(z)[ \, 1- \frac{3}{2} X'(z)^2 \, ]
\end{displaymath}
The elastic strain writes $\epsilon (\mathsf{x},z)=[\mathsf{x}-n(z)]/r(z)$ with $n(z)$ the neutral line. The coordinate system $(\mathsf{x},\mathsf{y},\mathsf{z})$ is attached to the cantilever, and is deduced from $(x,y,z)$ by a rotation of angle $\theta(z)$ around $y$. The $z$ coordinate runs from 0 to $z_{max} (F_{norm})$. By definition $\tan \theta(z) = X'(z)$ and $\vec{z}.\vec{\mathsf{z_m}} = \cos \theta(z_{max})$ ($\vec{\mathsf{z_m}}$ being the vector $\vec{\mathsf{z}}$ at $z=z_{max}$).
The (symmetric) stress tensor $\underline{\sigma}$ has four nonzero components $\sigma_{\mathsf{x}\mathsf{x}}$, $\sigma_{\mathsf{z}\mathsf{z}}$ and $\sigma_{\mathsf{z}\mathsf{x},\mathsf{x}\mathsf{z}}$. \\
Solving the problem is a tedious task which we will not reproduce here. It requires the usual boundary conditions ($X=0$, $dX/dz=0$ at $z=0$) plus the equation arising from the fixed length of the neutral line $n(z)$. The distortion $X(z)$  of the foot writes:
\begin{eqnarray*}
&X(z)& \!\!\! =  h    \Biggl[ \\
& \!\!\!\!\!\!\!\! & \!\!\!\!\!\!\!\!\!\!\!\!\!\!\!\!\!\!\!\!  \left(\frac{\alpha_{0}(z/h) (-2 +  \alpha_{0} (z/h)(1+ \alpha_{0}) ) - 2\,(1- \alpha_{0} (z/h) ) \ln (1-\alpha_{0} (z/h)) }{2 \, \alpha_{0}^3 \left(1 -  \alpha_{0} (z/h)\right) } \right) F_{norm} \\
&\!\!\!\!\!\!\!\!+ & \!\!\!\!\!\!\!\!\!\!\!\!\! \frac{e(0)}{h}\left(\frac{3 (z/h)^2}{8}-\frac{(z/h)^3}{12}+\alpha_{0} \left(\frac{7 (z/h)^2}{24}+\frac{11 (z/h)^3}{72}-\frac{ (z/h)^4}{16} \right) \right) F_{norm}^2 \\
& \!\!\!\!\!\!\!\!+ & \!\!\!\!\!\!\!\!\!\!\!\!\! \Biggl( -\frac{(z/h)^2}{15}+\frac{(z/h)^3}{45}+\frac{(z/h)^4}{8}-\frac{3(z/h)^5}{20}+\frac{(z/h)^6}{16}-\frac{(z/h)^7}{112}\\
& \!\!\!\!\!\!\!\!\!\!\!\!\!\!\!\!\!\!\!\!\!\!\!\!\!\!\!\!\!\!\!\!\!\!\!\!\!\!\!\!\!\!\!\!\!\!\!\!\!\!\!\!+ & \!\!\!\!\!\!\!\!\!\!\!\!\!\!\!\!\!\!\!\!\!\!\!\!\!\!\!\!\!\!\!\!\!\!\! \alpha_{0} \left(-\frac{13 (z/h)^2}{120}-\frac{23 (z/h)^3}{360}+\frac{ (z/h)^4}{20}+\frac{9 (z/h)^5}{20} -\frac{5 (z/h)^6}{8}+\frac{33 (z/h)^7}{112}-\frac{3 (z/h)^8}{64} \right)\Biggr) F_{norm}^3\Biggr]
\end{eqnarray*}
where the second and third order terms in $F_{norm}$ have been developed at first order in $\alpha_{0}=\alpha h / e(0)$ and $e(0)/h$. The maximum coordinate $z_{max}(F_{norm})$ is:
\begin{eqnarray*}
&z_{max}&= h \Biggl[1- \frac{e(0)}{h} \left( \frac{\alpha_{0} + \ln (1-\alpha_{0}) }{2\, \alpha_{0}^2} \right) F_{norm}
- \left( \frac{1}{15} +  \frac{13 \alpha_{0}}{120}  \right) F_{norm}^2 \\
 &&- \frac{e(0)}{h}\left( \frac{27}{160} +  \frac{277 \alpha_{0}}{720}   \right) F_{norm}^3 \Biggr]
\end{eqnarray*}
with the same expansions as above. We thus obtain for the maximum deflection $X_m(F_{norm})=X(z=z_{max})$:
\begin{eqnarray*}
&X_{m}&=h \Biggl[ \frac{1}{3} \left( -3 \frac{\alpha_{0}(2+\alpha_{0})+2\ln (1-\alpha_{0}) }{2\, \alpha_{0}^3} \right) F_{norm}
+ \frac{e(0)}{h} \left(  \frac{3}{8} +  \frac{\alpha_{0}}{2} \right) F_{norm}^2 \\
&& -\left(  \frac{4}{105} +  \frac{229 \alpha_{0}}{2240} \right) F_{norm}^3 \Biggr]
\end{eqnarray*} 
and the inverse function $F_{norm}(X_m)$: 
\begin{eqnarray*}
&F_{norm}&= 3 \left( \frac{2\, \alpha_{0}^3}{-3(\alpha_{0}(2+\alpha_{0})+2\ln (1-\alpha_{0})) } \right) \left( \frac{X_m}{h}\right) \\
&& + \frac{e(0)}{h} \left( - \frac{81}{8} +  \frac{297 \alpha_{0}}{32} \right) \left( \frac{X_m}{h} \right)^2    \\
&& +\left(  \frac{108}{35}- \frac{2187 \alpha_{0}}{2240} \right) \left( \frac{X_m}{h} \right)^3
\end{eqnarray*}
the first term being exact, and the others being again expansions in $\alpha_{0}=\alpha h / e(0)$ and $e(0)/h$. 

\vspace{0.15in}
We then inject the above expression of the distorted foot in the Rayleigh method presented in section \ref{rayleigh}. 
One gets for the non-linear spring constant $k_{v}$ and vibrating mass $m_{v}$:
\begin{eqnarray*}
&&k_{v}  =  k_{v0}  \Biggl[ 1 - \frac{e(0)}{h}\left(  \frac{ 9-3  \alpha_{0} }{4}\right) \left( \frac{X_m}{h} \right) + \left(  \frac{  288 + 243  \alpha_{0} }{1120}\right) \left( \frac{X_m}{h} \right)^2 \Biggr] , \\
&& \!\!\!\!\!\!\!\!\!\!\!\!\!\!\!\!\!\!\! m_{v}   =   m_{v0} \Biggl[ 1 - \frac{e(0)}{h}\left( \frac{ 146190 - 185447 \alpha_{0} }{58080} \right)\left(\frac{X_m}{h} \right) - \left( \frac{ 2118666 + 64357  \alpha_{0} }{1490720} \right) \left( \frac{X_m}{h} \right)^2 \Biggr]
\end{eqnarray*}
where we expressed the non-linearity in terms of the maximum displacement $X_m$. The first expression corresponds to a non-linearity in the {\em restoring force} (potential energy), while the other is an {\em inertia} non-linearity (kinetic energy). We find $k_{v0}=\frac{3 \, E_z I_z(0)}{h^3} \, (1 -\frac{3}{4} \alpha_0)$ and $m_{v0}= \rho e(0) e'  h \frac{33}{140} \, (1- \frac{133}{132}\alpha_0)$, all expressions at first order in $\alpha_{0}=\alpha h / e(0)$ and $e(0)/h$. \\
The linear contribution of the metal layer can be incorporated by the same changes as in section \ref{feetstatic} on $\rho$ (mass) and $I_z(0)$ (elasticity), applied only on the first order term. Higher order corrections are neglected$^($\footnote{This procedure neglects a (small) thickness gradient term $\alpha$ appearing in the metal-dependent inertia calculation. Higher order terms in $X_m$ require a full treatment of the metal-plus-Silicon beam, which is out of the scope of this paper.}$^)$. 

\vspace{0.15in}
The signal we measure has also a non-linear signature through the $\vec{z}.\vec{\mathsf{z_m}}$ term:
\begin{equation}
\vec{z}.\vec{\mathsf{z_m}}= 1- \left(\frac{ 9 }{8}+\frac{ 9\alpha_{0} }{16} \right) \left( \frac{X_m}{h} \right)^2 .
\label{signal}
\end{equation}
These expressions are used in section \ref{landau} to derive the dynamical equation of the system. The aim is to describe the (driven) first harmonic non-linear resonance, which is a general and important issue for MEM devices\cite{nonlinMEM}. The calculations are quantitatively compared to experiments in section \ref{datanalys}.

\subsection{Friction: dissipative and reactive terms}
\label{dissipe}

Friction is an essential component of dynamics since it allows to relax energy to a thermal bath when a substance's constituents are put into motion; effectively in any object made of {\em real} materials, such (complex) friction processes are present. Defining the nature of these mechanisms is another issue, and various material-dependent microscopic models can be found in the literature (see discussion of section \ref{themetal}). \\
However, beyond the nature of the mechanisms, friction models can be cast into constitutive equations governed by material-dependent coefficients, which we will extract from experiments in part \ref{datanalys}. Three main categories can be distinguished:
\begin{itemize}
\item {\em Static friction}, also called Coulomb friction is the force experienced by two solids at (relative) rest in contact, like the metal layer deposited on the Silicon feet. When this force is smaller than a threshold $\tau_{static}$, there is no slippage at the interface (see discussion in part \ref{datanalys}).
\item {\em Dry friction}, or sliding friction is constant in modulus and opposed to the displacement $\pm \tau_{dry}$. This is typically the force experienced by two solids in contact which have overrun the static friction limit. This type of forces shall not be discussed in this article\cite{dry}.
\item {\em Dynamic} friction, or viscous friction, is encountered when a solid is moving in a fluid. The materials called solid and fluid can be real materials like our vibrating wire moving in $^4$He gas, or idealized materials like the fluid-like low energy excitations of a piece of matter containing them. Actually, any friction process in a solid proportional to the strain rate $\dot{\epsilon}$ will look like a viscous force (see section \ref{internal}). 
\end{itemize}
Linear viscous friction occurs in so-called {\em Newtonian fluids} (for instance $^4$He gas, see Eq. (\ref{totfluid}) section \ref{gas}). It decomposes into a damping force proportional to $\dot{x}$ (transferring energy from the solid to the fluid), and a reactive force proportional to $\ddot{x}$ (corresponding to a boundary layer of fluid put into motion by the solid). These forces in the (angular) frequency space $\omega$ write:
\begin{eqnarray*}
  dissipative\,f\!orce & = & -2\Lambda_1 (\omega) \, i \omega \, \tilde{x}(\omega) ,\\
  reactive\,f\!orce & = & +2\Lambda_2 (\omega)  \, \omega^2 \, \tilde{x}(\omega) 
\end{eqnarray*}
the tilde denoting the Fourier transform. The frequency dependence of the damping coefficients $\Lambda_{1,2}$ originates in the finite time required by the friction mechanism to take place. The canonical case is simply $\Lambda_{1,2}=\,$Cst corresponding to instantaneous response. $\Lambda_{1,2}$ can present a peaked structure as different mechanisms with different characteristic timescales are resonantly coupled to the oscillator. However, in general $\Lambda_{1,2} (\omega)$ are rather slow functions of $\omega$, and are considered as constant over the width of the studied resonance (see following sections).

\vspace{0.15in}
Subclasses of dynamic friction can be distinguished with the non-linear extensions of viscous friction.  \\
The most obvious non-linear extension considers a friction proportional to $\dot{x}^2$. This is typically the drag force arising at large Reynolds numbers in a fluid. In this paper we restrict the $^4$He study in part \ref{datanalys} to small velocities, thus this non-linearity will not be further addressed. \\
The other non-linear extension considers that the friction mechanism {\em itself} depends on the strain state of the material. This {\em non-Newtonian fluid} behavior is for instance the case of viscoelastic fluids (or Maxwell materials). Various ways of extending, at first order, the above equations can be envisaged. For reasons which are presented in the experimental sections, we discuss a model describing a dilatant material, or {\em shear thickening fluid}. In this model, the friction terms $\Lambda_{1,2}$ arise from a viscosity $\nu$ which depends on 
the shear stress rate $\left| \dot{\sigma} \right|$ in the material:
\begin{equation}
\nu = \nu_0 + \nu_1 \, \left| \dot{\sigma} \right| \label{nonlinnu}
\end{equation}
at first order. For a (small) sinusoidal solicitation, the stress will (mainly) have a harmonic component $\sigma(t)=\left|\tilde{\sigma}(\omega)\right|\, \cos [\omega t + \varphi(\omega)]$. Finding the first harmonic response of the oscillator requires only$^($\footnote{Higher order modes will have signatures comparable to an $\dot{x}^2$ non-linearity, which again, is not of the type addressed in the experimental sections. }$^)$ the zero frequency mode of $\nu$ in Eq. (\ref{nonlinnu}). We thus write:
\begin{displaymath}
\nu = \nu_0 \left( 1+ \frac{2}{\pi}\frac{\nu_1}{\nu_0} \, \omega \left| \tilde{\sigma} (\omega) \right| \right)
\end{displaymath}
with $\tilde{\sigma} (\omega)$ the Fourier transform of $\sigma(t)$. Since the stress $\sigma$ is proportional to the displacement $x$, adapting the fluid friction model (sections \ref{internal} and \ref{gas}) brings:
\begin{eqnarray*}
\Lambda_1 (\omega) & \rightarrow &  \Lambda_1 (\omega)   + \Lambda_{1}'(\omega) \, \left| \tilde{x} (\omega) \right|  , \\
\Lambda_2 (\omega) & \rightarrow &  \Lambda_2 (\omega)   + \Lambda_{2}'(\omega) \, \left| \tilde{x} (\omega) \right|  .
\end{eqnarray*}
For a narrow resonance, the coefficients $\Lambda_{1,2}'(\omega)$ can again be chosen to be constants.\\
This formalism is derived in the next sections addressing the various dissipation terms encountered in part \ref{datanalys}.

\subsubsection{Internal friction}
\label{internal}

We call {\em internal friction} all friction processes occurring inside the oscillator. It is made of low Boron-doped monocrystalline Silicon, covered with a thin (polycrystalline) metal layer. \\
For bare Silicon, the quality factor $Q$ measured both by flexural or torsional oscillations scales with the size of the MEM device\cite{Qfact}. Our Silicon vibrating wire has dimensions comparable to those used in the Cornell work\cite{ParpiaQ}, which would imply a $Q$ factor in the range of a few $10^6$ for Kelvin temperatures. Furthermore, bare (doped) Silicon displays in the same range of temperatures a magnetic dependence on both the resonance frequency and the dissipation\cite{ParpiaQfield}. In our work with metal-coated Silicon, the highest $Q$ is smaller than $0.5\,10^6$, without any measurable magnetic field dependency of the oscillator's properties above $1.5\,$K. We thus conclude that the friction mechanisms are always, in our case, dominated by the metal layer.

\vspace{0.15in}
We take the shear thickening fluid model of the previous section as a description for the metal layer. Obviously, only the regions experiencing distortions will participate to friction (namely the two feet). At the Silicon surface, the additional stress induced by the applied force $F(t)$ is only axial. This $\sigma_{zz}$ component can be evaluated by use of the dynamical shape $X(z,t)$ obtained with the Rayleigh method, substituting $X_m \rightarrow X_m(t)$ in the static expression of one foot:
\begin{displaymath}
\left|\sigma_{zz}(x=0,z,t)\right|  =   E_z \frac{ \left| X_m(t)\right|}{ h } \frac{e(0)}{h} \frac{3}{\left( 1+ \frac{3}{4} \alpha_0  \right) }  \frac{1}{2}  \left[1- \frac{z}{h} + 3\frac{z}{h}\left( 1- \frac{z}{h} \right)\alpha_0 \right] 
\end{displaymath}
at the lowest orders (in $X_m$ and $\alpha_0$, the thickness gradient already introduced), and neglecting the static stress in the materials. \\
The ideal force per unit length, Eq. (\ref{fluideq}), of the next section gets modified in the following way by the stress-dependent viscosity, Eq. (\ref{nonlinnu}):
\begin{displaymath}
 \kappa + i \kappa' \rightarrow \kappa(\nu_0) + i \kappa'(\nu_0) + \left(\frac{d \kappa}{d \nu}(\nu_0) + i \frac{d \kappa'}{d \nu}(\nu_0)\right)\nu_1 \, \left| \dot{\sigma_{zz}} \right|
\end{displaymath}
at first order ($\kappa$ and $\kappa'$ being the Stokes' functions). Integrating the friction along the feet finally produces for the lowest order term:
\begin{eqnarray*}
\!\!\!\!\! total\,f\!orce & = & - \, i \omega \, \rho_{fluid} \,2  \Biggl[ \left(\frac{3}{8}-\frac{9 \alpha_0}{160} \right) \, e'e_M  h  \,\left( \kappa + i \kappa' \right) \\
& + &  \left( \frac{1}{10}+ \frac{3 \alpha_0}{40}\right)\frac{3}{\pi} \omega \, e' e_M  h \, \frac{e(0)}{h} \,E_z  \left( \frac{d \kappa}{d \nu} + i\frac{d \kappa}{d \nu}' \right) \nu_1 \! \left|\frac{\tilde{X}_m(\omega)}{h}\right| \, \Biggr]\,\, i\omega \, \tilde{X}_m(\omega)
\end{eqnarray*}
where, according to the previous section, we kept only the first Fourier component of the non-linear viscosity. Introducing real and imaginary parts allows to define easily the coefficients $\Lambda_{1,2}$ and $\Lambda'_{1,2}$. \\
Note that the above equation, although perfectly well defined, is nothing more than an {\em effective model}. As a matter of fact, $\rho_{fluid} \, \kappa$, $\rho_{fluid} \, \kappa'$ and $\rho_{fluid} \, \frac{d \kappa}{d \nu}\, \nu_1$, $\rho_{fluid} \, \frac{d \kappa'}{d \nu}\, \nu_1$ are four effective material-dependent parameters which have to be fit on experiments. The only physical statement behind this formalism is that the friction process in the solid is proportional to the strain rate $\dot{\epsilon}$ (or equivalently to the stress rate $\dot{\sigma}$) and can be further developed at the lowest orders in a series of $\left|\dot{\epsilon}\right|$ (equivalently $\left|\dot{\sigma}\right|$). Note also that the total force is proportional to the metal thickness $e_M$ (or more precisely to the volume $e'\,e_M \,h$). \\
Moreover, it is experimentally impossible to distinguish a change in the reactive part of the friction (through the $\Lambda_2$ term), from an opposite change in the spring constant of the oscillator (through the stored stress $S$, expressed by $u$ in the preceding sections), since both simply {\em shift} the resonance in the same direction (section \ref{harmo}). 
Our pragmatic approach in part \ref{datanalys} shall get round this difficulty. 

\vspace{0.15in}
The non-linear Rayleigh method naturally produces a non-linear global friction for the structure (of geometrical origin). We take the friction coefficients $\Lambda_{1,2}$ as constants (and proportional to the volume of the metal layer), and write $-2 \Lambda_1 \, \dot{x} (t)$ in the $t$ space for the dissipative part, and $-2 \Lambda_2 \, \ddot{x} (t)$ for the reactive part of the friction force. 
Using the expressions of section \ref{nonlin}, and keeping the lowest orders in $X_m$, we obtain for the structure after integration along the feet:
\begin{eqnarray}
 total\,f\!orce & = &  - 2\, \Lambda_{1} \left[ 1 + \frac{e(0)}{h}\left(\frac{1}{2}+\frac{7 \alpha_0}{80}\right) \frac{X_m}{h} -\left(\frac{11}{35}-\frac{331 \alpha_0}{33600}\right) \left(\frac{X_m}{h}\right)^2\right] \dot{X}_m \nonumber \\
  &- &  2 \, \Lambda_{2} \left[ 1 + \frac{e(0)}{h}\left(\frac{1}{2}+\frac{7 \alpha_0}{80}\right) \frac{X_m}{h} -\left(\frac{11}{35}-\frac{331 \alpha_0}{33600}\right) \left(\frac{X_m}{h}\right)^2\right] \ddot{X}_m \nonumber \\
  &- & 2 \, \frac{\Lambda_{2}}{h}  \left[\!- \frac{e(0)}{h}\left(\frac{1}{4}-\frac{3 \alpha_0}{20} \right) - \left(\frac{351}{280}+\frac{2421 \alpha_0}{5600 }\right) \frac{X_m}{h} \right]  \dot{X}_m^2 \label{nonlindissipe}
\end{eqnarray}
were only the first order in $\alpha_0$ and $e(0)/h$ have been kept. The reactive and dissipative parts are now functions of $X_m (t)$. Note that an abnormal term proportional to $\dot{X}_m (t)^2$ has appeared, which is formally of the same type as a viscous $\dot{x}^2$ non-linearity. \\
Both material-dependent, and geometry-dependent friction non-linearities are considered in the experimental study.

\subsubsection{Fluid friction}
\label{gas}

An additional fluid friction appears on the structure when it is immersed in gaseous or liquid $^4$He at low temperatures. The dynamics of a straight infinite cylinder oscillating transversely in a Newtonian fluid was first solved by Stokes\cite{stokes}. The force per unit length exerted by the fluid on the solid writes:
\begin{equation}
fluid\,f\!orce\,per\,length = - \, i \omega \,\rho_{fluid} \, S_{beam} \,\left( \kappa + i \kappa' \right)\,\, i\omega \, \tilde{x}(\omega) \label{fluideq}
\end{equation}
in the (angular) frequency space. $S_{beam}$ is the section of the moving cylinder (or volume per unit length), and $i\omega \, \tilde{x}(\omega)$ its speed. $\rho_{fluid}$ is the mass density of the surrounding gas or liquid. \\
The Stokes' functions $\kappa$ and $\kappa'$ are obtained under the assumptions\cite{landaufluid} of linearized Navier-Sokes equations (low speed or small displacements$^{(}$\footnote{Comparing $\delta$ (viscous length) to $r$ (wire radius) brings two regimes with two different criteria: for $\delta \gg r$, the Reynolds number defined by $Re= (x \, \omega) (2 r) \rho_{fluid} / \nu$ has to be small $Re \ll 1$ (with $x$ the displacement of the body oscillating at $\omega$), and for  $\delta \ll r$ one needs only $x \ll 2 r$.}$^{)}$), incompressibility of the fluid$^($\footnote{In the ideal fluid, the stress tensor generating the friction depends only linearly on the velocity gradient inside the fluid. The most general tensor satisfying the symmetries implies two positive scalars, $\nu$ (shear viscosity) and $\varsigma$ (second viscosity, associated to volume change). For an incompressible fluid, the contribution of $\varsigma$ vanishes. But even in a compressible flow (as in a gas), incompressibility can be stated if the velocities are much smaller than the velocity of sound. However, in some peculiar cases $\varsigma$ and its determination is a matter of controversy.}$^)$, and perfect clamping of the boundary layer to the solid surface. They imply the definition of a length scale $\delta$:
\begin{displaymath}
\delta =  \sqrt{\frac{2 \, \nu }{\rho_{fluid} \, \omega}}
\end{displaymath}
called the viscous penetration depth. It corresponds to the typical length scale over which the fluid is put into motion. $\nu$ is here the (dynamic) viscosity of the fluid (while $\nu/\rho_{fluid}$ is the kinematic viscosity). The expression of the Stokes' functions in terms of Bessel functions$^($\footnote{For the whole article, the Fourier (series) transform convention corresponds to $e^{+ i \omega t}$ factors; in the literature, the above expressions are usually written with the opposite $e^{- i \omega t}$.}$^)$ is:
\begin{displaymath}
\kappa-1 + i \kappa'  = -\frac{4}{[1-i] \, r/\delta} \, \frac{J_{1} \left([1-i] \, r/\delta \,\right) - i \, Y_{1}\left([1-i] \, r/\delta \,\right) }{J_{0} \left([1-i] \, r/\delta \,\right) - i \, Y_{0}\left([1-i]  \, r/\delta _,\right) }
\end{displaymath}
with $r$ the radius of the cylinder. $\kappa$ corresponds to a reactive component, in phase with the acceleration, and $\kappa'$ to a dissipative component in phase with the speed of the moving object.\\
The fluid force Eq. (\ref{fluideq}) can be modified\cite{Sader} for an infinite 
rectangular beam of width $2\,r$ and thickness $e$ in the following way:
\begin{displaymath}
(\kappa + i \kappa') \rightarrow \,\frac{\pi}{4} \frac{2\,r}{e} \, \Omega(\omega) \, (\kappa + i \kappa')
\end{displaymath}
where $\Omega(\omega)=\Omega_r(\omega)- i \,\Omega_i(\omega)$ is a correction function. 
The section $S_{beam}=\pi\,r^2$ of the cylinder is then replaced by $S_{beam}=2 r \,e$. These changes simply state that the dominant lengthscale for the hydrodynamic flow is the transverse $2\,r$; the length $e$ has formally disappeared in the expression of the force, meaning that the flow around the object will not be too far from that around a cylinder of radius $r$. \\
Indeed, for a very thin beam$^($\footnote{Since we are talking of corrections of the order of 15$\,\%$, even in a square geometry where $2\,r \approx e$ the correction coefficient $\Omega$ reproduced here has to be rather good, especially in the limit $\delta \ll r$ (see part \ref{datanalys}).}$^)$, $\Omega(\omega)$ has been fit on numerical simulations\cite{Sader} and appears to modify Stokes' result by at most 15$\,$\%. It is expressed as a rational function of $\tau = \log_{10} 2 \, (r/\delta)^2$ having the correct asymptotic behavior:
\begin{eqnarray*}
\Omega_r(\omega) & = & (0.91324 - 0.48274 \tau + 0.46842 \tau^2 - 0.12886 \tau^3 + 
  0.044055 \tau^4 \\
 & - & 0.0035117 \tau^5 + 0.00069085 \tau^6)/(1 - 0.56964 \tau + 0.48690 \tau^2 \\
 &-& 0.13444 \tau^3 +   0.045155 \tau^4 - 0.0035862 \tau^5 + 0.00069085 \tau^6) , \\
\Omega_i(\omega) & = & (-0.024134 - 0.029256 \tau + 0.016294 \tau^2 - 0.00010961 \tau^3 \\
& + &  0.000064577 \tau^4 - 0.000044510 \tau^5)/(1 - 0.59702 \tau + 0.55182 \tau^2 \\
&- & 0.18357 \tau^3 + 0.079156 \tau^4 -   0.014369 \tau^5 + 0.0028361 \tau^6)
\end{eqnarray*}
which are considered to be accurate within $0.1\,\%$. 
In the above equations, the quantity $r$ is the half-width of the oscillating structure considered, that is $a/2$ for the paddle and $e'/2$ for the feet. \\
Finite slippage at the boundary layer can be incorporated in the model\cite{CHH,pariaslip,bowley,slippapergas}. It is expected to be of importance in dilute fluids (like low pressure $^4$He gas, or cold quantum liquids), and/or for smooth surfaces (like our Silicon devices). One introduces then a slippage parameter $\beta$ by substituting:
\begin{displaymath}
\kappa - 1 + i \kappa' \rightarrow  \frac{1}{(\kappa-1 + i \kappa')^{-1} + \frac{1}{2} i \, \beta \, (r/\delta)^2} .
\end{displaymath}
Physically $\beta$ is related to the mean free path $l_{m\!f\!p}$ of the particles in the fluid (or quasi-particles in a quantum fluid). As $l_{m\!f\!p}$ becomes comparable or larger than $r$, an hydrodynamic treatment near the solid is not strictly valid. However, one can take into account (at first order) the finite mean free path by modifying the boundary condition
on the moving object. The gradient of the radial velocity in the vicinity of the solid defines a length scale $\xi$, proportional to $l_{m\!f\!p}$,  over which it would extrapolate to zero {\em beyond} the surface. The dependence of $\beta$ on this so-called slip length $\xi$ is a function of the geometry\cite{bowley}; for a flat surface (which is our limiting model) $\beta=\xi/(\xi + r)$ while for a cylinder $\beta=\xi/(2 \, \xi + r)$. Note that one needs $l_{m\!f\!p}/\delta \ll 1$ in order to apply hydrodynamics in the bulk of the fluid.

\vspace{0.15in}
The total force acting on the oscillator is obtained by integrating the force per unit length along the structure (feet plus paddle). In order to describe the dynamical shape $X(z,t)$ we apply again the Rayleigh method, and substitute $X_m \rightarrow X_m(t)$ in the static expression of one foot. Limiting the calculation to the linear term, and neglecting the static stress in the material, Eq. (\ref{fluideq}) produces:
\vspace{0.5in}
\begin{eqnarray}
\!\!\!\!\!\!\!\!\!\! \!\!\!\!\! \!\!\!\!\!\!\!\!\!\!  total\,f\!orce &=& - \, i \omega \,\rho_{fluid} \, \Biggl[ 2 \left(\frac{3}{8}-\frac{9 \alpha_0}{160} \right) e'\,e(0)\,h  \, \left( \kappa + i \kappa' \right)\nonumber \\
&+& a \, e(h)\, l \, \left( \kappa + i \kappa' \right)\,\Biggr] \,\, i\omega \, \tilde{X}_m(\omega) \label{totfluid}
\end{eqnarray}
where the above corrections have to be applied to $\left( \kappa + i \kappa' \right)$. Note that the first $\left( \kappa + i \kappa' \right)$ is calculated using $r=e'/2$ while the last one uses $r=a/2$. $\tilde{X}_m(\omega)$ is the Fourier transform of $X_m(t)$ and $e(z)$ is the $z$-position (linearly) dependent thickness of one foot ($\alpha_0$ being the thickness gradient term already introduced; the expression is an expansion at first order). The assumption behind this treatment is obviously that the viscous length $\delta$ is small compared to the lengths of the beams ($\delta \ll l$, $\delta \ll h$), allowing to treat them as infinite.
Inserting real and imaginary parts in Eq. (\ref{totfluid}) enables to define easily the coefficients $\Lambda_{1,2}$ presented in the introductory section. They are used in sections \ref{harmonic} (linear) and \ref{landau} (non-linear) in order to produce the equivalent one dimensional equation of motion of the whole structure.

\subsection{Non-linear first harmonic solution}
\label{landau}

The effective one dimensional non-linear dynamics equation, replacing the linear Eq. (\ref{equa}) completed with the viscous terms, can be written in the most general form:
\begin{eqnarray}
  & & \!\!\!\!\!\!\!\!\!\!\!\!\!\!\! \left[ \Delta m_v + 2 m_{v0} \left(1 +m_{v1} X_m + m_{v2} X_m^2 \right) \right] \ddot{X}_m   \nonumber \\
&+& 2\, \left[\Delta \Lambda_2 + 2\Lambda_{2} \left( 1+ \Lambda_{21} X_m + \Lambda_{22} X_m^2 \right) + 2\Lambda_{2}' \left|\tilde{X}_m (\omega) \right|  \, \right]\ddot{X}_m   \nonumber \\
&+& 2\, \left[\Delta \Lambda_1 + 2\Lambda_{1} \left( 1+ \Lambda_{11} X_m + \Lambda_{12} X_m^2  \right) + 2\Lambda_{1}' \left|\tilde{X}_m (\omega) \right| \, \right]\dot{X}_m \nonumber \\
&+&  \left( \Gamma_0 + \Gamma_1 X_m \right) \dot{X}_m^2 \nonumber \\
&+& \left[ \Delta k_v + 2 k_{v0} \left( 1+ k_{v1} X_m + k_{v2} X_m^2\right) \right]X_m \,\,\,\,\,\,\, = F(t) \label{dynamic}
\end{eqnarray}
keeping third order terms in $X_m$ (the displacement of the end of the structure, dots being time derivatives). The parameters $m_{vi}$, $k_{vi}$ are respectively the non-linear mass and spring coefficients ($i=1,2$). The $\Lambda_{2i}$, $\Lambda_{1i}$ are the intrinsic non-linear reactive and dissipative coefficients. These non-linear terms associated to each foot are of {\em geometrical} origin, and have been evaluated previously using the Rayleigh method of section \ref{nonlin}. In our description, $\Gamma_0$ and $\Gamma_1$ are also produced by the geometrical non-linearity, but in the most general context they could include a non-linear drag force (due to an $\dot{x}^2$ contribution to the damping).\\
$m_{v0}$ and $k_{v0}$ are the (linear) vibrating mass and spring of one foot of the structure.
The parameters $\Delta m_v$ and $\Delta k_v $ represent the effect of the metal layer on the (linear) mass and spring (through $\rho_M$ and $E_M$). Note that formally the static stress $S$ is also included in $\Delta k_{v}$ (through a term $u \ll 1$ according to the preceding sections). We incorporate in $\Delta m_v$ the mass of the paddle, too. The coefficients $\Lambda_{1}'$ and $\Lambda_{2}'$ correspond to the metal stress-rate dependent friction terms, arising from Eq. (\ref{nonlinnu}), $\tilde{X}_m (\omega)$ being the Fourier transform of $X_m (t)$. \\
$\Lambda_1$ and $\Lambda_2$ are the intrinsic friction coefficients of one foot. 
Finally, $\Delta \Lambda_1$ and $\Delta \Lambda_2$ are the fluid (linear) damping contributions when the structure is immersed for instance in $^4$He gas, Eq. (\ref{totfluid}).

\vspace{0.15in}
In order to solve the above equation Eq. (\ref{dynamic}) for forced oscillations $F(t)=F_{ac} \cos ( \omega t)$, we extend Landau's\cite{landaumeca} non-linear technique. We postulate for the solution the form:
\begin{displaymath}
X_m (t) = \sum_{n=0}^{+\infty} a^c_{n}(\omega) \cos (n\,\omega t) + \sum_{n=1}^{+\infty} a^s_{n}(\omega) \sin (n\,\omega t)
\end{displaymath}
and seek only the first harmonics$^($\footnote{One has to be careful not confusing the {\em normal modes} of the structure $\omega_{0}^{(j)}$, with the {\em harmonics} of each mode $n\,\omega_{0}^{(j)}$ ($n>0$ integer) resulting from the non-linear terms.}$^)$ $n \leq 1$. By definition $\tilde{X}_m (\omega)= X\!P_m - i X\!Q_m$ in the Fourier $\omega$ space, $X\!P_m= a^c_{1}(\omega)$, $X\!Q_m= a^s_{1}(\omega)$, and we define a static deflection $X_0=a^c_{0}(\omega)$. Replacing the above expression in Eq. (\ref{dynamic}), and assuming that the higher orders amplitudes $a^{c,s}_{n>1}(\omega)$ are all negligible, we obtain:
\begin{eqnarray}
\!\!\!\!\!\!\!\!\!\!\!\!\!\!\!\!\!\!\!\!\!\!\!\!\! X\!P_m &=& \frac{F_{ac}}{2k_{v0}+\Delta k_v} \frac{(\omega_r/\omega_{0})^2-(\omega/\omega_{0})^2- 2 \delta \omega_0 \, \omega /(\omega_0 )^2}{\left[(\omega_r/\omega_{0})^2-(\omega/\omega_0)^2- 2\delta \omega_0 \,\omega /(\omega_0)^2\right]^2+ \left[\Delta \omega \, \omega/(\omega_0)^2\right]^2}  , \label{nonlinphase}\\
\!\!\!\!\!\!\!\!\!\!\!\!\!\!\!\!\!\!\!\!\!\!\!\!\! X\!Q_m &=& \frac{F_{ac}}{2k_{v0}+\Delta k_v} \frac{\Delta \omega \, \omega/(\omega_0)^2}{\left[(\omega_r/\omega_{0})^2-(\omega/\omega_0)^2- 2\delta \omega_0 \,\omega /(\omega_0)^2\right]^2+ \left[\Delta \omega \, \omega/(\omega_0)^2\right]^2} . \label{nonlinquad}
\end{eqnarray}
The resonance equations have formally the same structure as Eq. (\ref{phase}) and (\ref{quad}), and we shall call it a {\em modified Lorentzian}. We introduced:
\begin{eqnarray*}
\!\!\!\!\!  \omega_r^2 = \omega_0^2 + 2 b_0 \,\omega_0\, \left|\tilde{X}_m (\omega)\right|^2 & i.e. & \omega_r \approx \omega_0 + b_0 \, \left|\tilde{X}_m (\omega)\right|^2 ,\\
\Delta \omega &=& \Delta \omega_0 + b_1 \, \left|\tilde{X}_m (\omega)\right|^2 ,\\
X_0    & = & b_2 \, \left|\tilde{X}_m (\omega)\right|^2 
\end{eqnarray*}
which are now functions of $\left|\tilde{X}_m (\omega)\right|^2= X\!P_m^2 + X\!Q_m^2 $. The linear definitions are unchanged, and produce:
\begin{eqnarray*}
\omega_0 &=& \sqrt{\frac{\Delta k_v + 2 k_{v0}}{\Delta m_v + 2 m_{v0}}} ,\\
\delta \omega_0 & = & \frac{\omega \left( \Delta \Lambda_2 + 2 \Lambda_2 + 2\Lambda_{2}' \left|\tilde{X}_m (\omega) \right| \, \right) }{\Delta m_v + 2 m_{v0}} ,\\
\Delta \omega_0  &= & \frac{2 \left( \Delta \Lambda_1 + 2 \Lambda_1 + 2\Lambda_{1}' \left|\tilde{X}_m (\omega) \right| \,\right)}{ \Delta m_v + 2 m_{v0} } 
\end{eqnarray*}
with the metal's friction non-linearity incorporated. $b_0$, $b_1$ and $b_2$ are the "experimental" non-linear coefficients, in the sense that only these three are required in order to describe the harmonic displacement $\tilde{X}_m (\omega)$. They are expressed in terms of the primarily defined coefficients:
\begin{eqnarray*}
\!\!\!\!\!\!\!\!\!\!\!\!\!\!\!\!\!\!\!\!\!\!\!\!\!\!\!\!\!\!\!\!\!\! b_0 & = & \frac{\omega_0}{2} \, \Biggl[ \, \frac{3}{4} \frac{2  k_{v0}}{\Delta  k_{v}+ 2  k_{v0}} k_{v2} \\
&+& \left( \left( \frac{2 \Lambda_2}{\Delta  m_{v}+ 2  m_{v0}}\Lambda_{21}+ \frac{1}{2}\frac{2  m_{v0}}{\Delta  m_{v}+ 2  m_{v0}} m_{v1} \right) \left(\frac{\omega}{\omega_0}\right)^2 - \frac{2  k_{v0}}{\Delta  k_{v}+ 2  k_{v0}} k_{v1} \right) \times \\
& & \!\!\! \left(\frac{2  k_{v0}}{\Delta  k_{v}+ 2  k_{v0}} k_{v1}+ \left(\frac{\omega}{\omega_0}\right)^2 \!\! \left( \frac{\Gamma_0}{\Delta  m_{v}+ 2  m_{v0}} - \frac{2  m_{v0}}{\Delta  m_{v}+ 2  m_{v0}} m_{v1} - 2 \frac{2 \Lambda_2}{\Delta  m_{v}+ 2  m_{v0}}\Lambda_{21} \right)\! \right) \\
& - & \left(\frac{\omega}{\omega_0}\right)^2 \!\! \left( \frac{3}{4} \frac{2 m_{v0}}{\Delta  m_{v}+ 2  m_{v0}} m_{v2} -\frac{1}{4}\frac{\Gamma_1}{\Delta  m_{v}+ 2  m_{v0}} + \frac{3}{2} \frac{2 \Lambda_2}{\Delta  m_{v}+ 2  m_{v0}}\Lambda_{22}  \right) \Biggr] ,\\
\!\!\!\!\!\!\!\!\!\! b_1 &=& \frac{2 \Lambda_1}{\Delta  m_{v}+ 2  m_{v0}} \Biggl[ \frac{1}{2} \Lambda_{12} - \Lambda_{11} \Biggl( \frac{2  k_{v0}}{\Delta  k_{v}+ 2  k_{v0}} k_{v1} \\
&+& \left(\frac{\omega}{\omega_0}\right)^2 \!\! \left( \frac{\Gamma_0}{\Delta  m_{v}+ 2  m_{v0}} - \frac{2  m_{v0}}{\Delta  m_{v}+ 2  m_{v0}} m_{v1} - 2 \frac{}{}  \frac{2 \Lambda_2}{\Delta  m_{v}+ 2  m_{v0}}\Lambda_{21} \right)\Biggr)\Biggr] ,\\
\!\!\!\!\!\!\!\!\!\! b_2 &=& - \frac{1}{2} \Biggl[ \frac{2  k_{v0}}{\Delta  k_{v}+ 2  k_{v0}} k_{v1} \\
&+& \left(\frac{\omega}{\omega_0}\right)^2 \!\! \left( \frac{\Gamma_0}{\Delta  m_{v}+ 2  m_{v0}} - \frac{2  m_{v0}}{\Delta  m_{v}+ 2  m_{v0}} m_{v1} - 2 \frac{}{}  \frac{2 \Lambda_2}{\Delta  m_{v}+ 2  m_{v0}}\Lambda_{21} \right) \Biggr]
\end{eqnarray*}
Of course, we recover the Lorentz linear case when $b_i =0$.
The resonance frequency is now defined as $\omega_{res}= \omega_r -\delta \omega_0 $.

\vspace{0.15in}
To conclude the modeling, we have to recall that the signal $V$ we measure is also non-linear. Consider a general detection scheme verifying:
\begin{equation}
V = V_0 \, \dot{X}_m \left[ 1 + V_1 \, X_m + V_2 \, (X_m)^2 \right] . \label{nonlinsig}
\end{equation}
Solving with the above technique brings a renormalization of the cut flux (using Fourier notations):
\begin{equation}
\tilde{V} (\omega) = B l \, i \omega \, \tilde{X}_m (\omega) \left[ 1 +  b_3 \, \left|\tilde{X}_m (\omega)\right|^2 \right]
\end{equation}
which is again a second order correction involving $\left|\tilde{X}_m (\omega)\right|^2$. It introduces the last "experimental" non-linear parameter $b_3$:
\begin{eqnarray*} 
\!\!\!\!\!\!\!\!\!\!\!\!\!\!\!\!\!\!\!\!\!\!\!\!\! b_3 & = &   \frac{1}{4} V_2 \\
&\!\!\!\!\!-\!\!\!\!\!&  \frac{1}{2} V_1 \left(\frac{2  k_{v0}}{\Delta  k_{v}+ 2  k_{v0}} k_{v1}+ \left(\frac{\omega}{\omega_0}\right)^2 \!\! \left( \frac{\Gamma_0}{\Delta  m_{v}+ 2  m_{v0}} - \frac{2  m_{v0}}{\Delta  m_{v}+ 2  m_{v0}} m_{v1} - 8 \frac{2 \Lambda_2}{\Delta  m_{v}+ 2  m_{v0}}\Lambda_{21} \right) \! \right) 
\end{eqnarray*}
defined from the primary non-linear coefficients of Eq. (\ref{nonlinsig}) and Eq. (\ref{dynamic}).
In our case, $V_1$ (with $V_1=0$) and $V_2$ are obtained through Eq. (\ref{signal}) in the Rayleigh approximation. 
 
\vspace{0.15in}
The non-linear resonance expressions Eq. (\ref{nonlinphase}) and Eq. (\ref{nonlinquad}) have to be solved numerically. However, they have some simple properties which are illustrated in part \ref{experiment} on the basis of our measurements. 

\section{RESULTS}
\label{datanalys}

The features presented in part \ref{experiment} are analyzed quantitatively in the following sections with the previous theoretical tools. The non-linearities and the materials are characterized. The damping due to an ideal fluid ($^4$He gas) is fit to theory without free parameters.

\subsection{Geometrical non-linearities}
\begin{figure}
\centerline{\includegraphics[width= 9. cm]{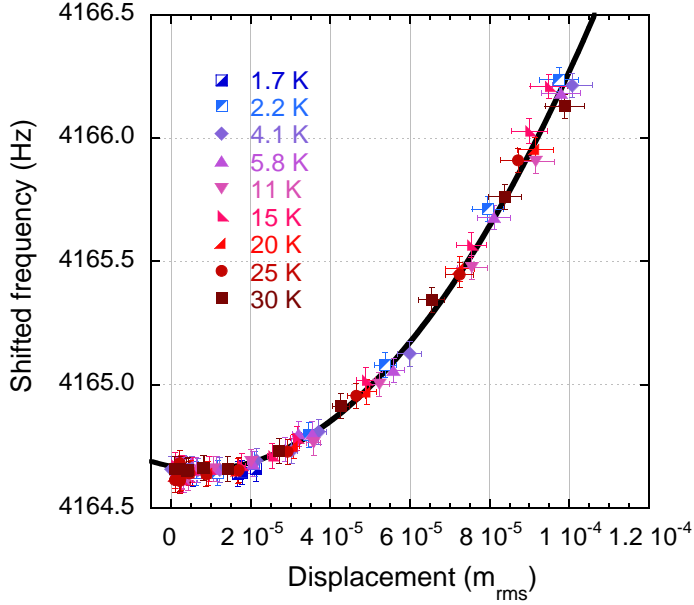}}
%
\caption{(Color online). Non-linear frequency shift as a function of the displacement $X_{max}$, measured in vacuum with a field around 10$\,$mT and various temperatures (E6 third metal deposition). The temperature contribution on the linear resonance frequency has been subtracted, in order to superimpose all the curves. Error bars on the $y$ axis are about $\pm50\,$mHz and about $\pm5\,$\% on the $x$ axis. The thick line is a fit (see text).}  
\label{tempnonlin}
\end{figure}

\begin{figure}
\centerline{\includegraphics[width= 9. cm]{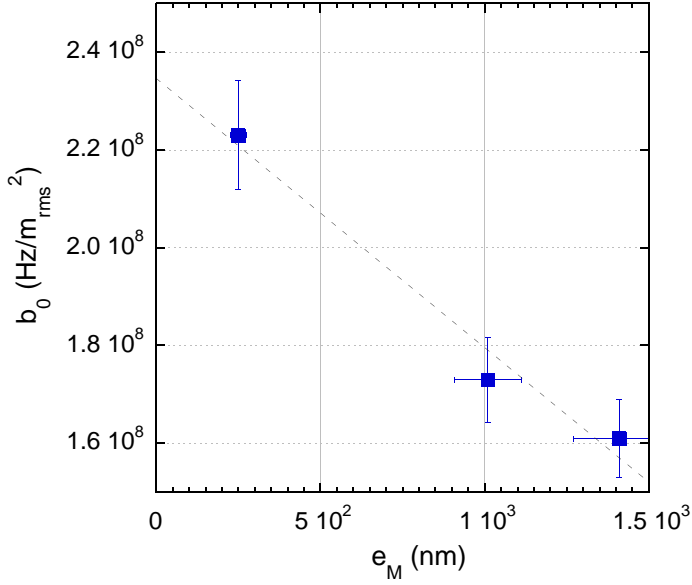}}
%
\caption{(Color online). Non-linear parameter $b_0$ extracted for sample E6 as a function of the metal thickness. The dashed line is a linear guide (see text).}  
\label{nonlinthick}
\end{figure}

In Fig. \ref{tempnonlin} the position of the resonance $f_{res}$ is presented as a function of displacement (similarly to Fig. \ref{freqnlin}), for various temperatures. The data were shifted, in order to remove the (linear) temperature-dependent effect presented in the next section. All measurements were taken in vacuum with fields around 10$\,$mT. As a result, we see that the non-linear quadratic shift is {\em independent of temperature}, from basically 1.5$\,$K to 30$\,$K. The data presented are those of sample E6 for the third metal deposition, but the conclusion holds regardless of the metal quantity. This confirms the geometrical origin for this effect. On the contrary, standard vibrating wires display a {\em temperature-dependent} non-linear coefficient\cite{pobellnonlin}, function of the metals used (typically NbTi, Tantalum). 
The slight decrease of the resonance frequency at small drives is discussed in the next section.\\
In Fig. \ref{nonlinthick} we show the dependency of the $b_0$ parameter to the metal thickness. As expected from section \ref{landau}, increasing the mass of the resonator makes the non-linear parameter decrease linearly.

In Tab. \ref{NonlinSamples} we give a summary of the geometrical dimensions and the non-linear parameters obtained from experiment and theory for two samples studied in this article. The (linear) properties $f_{res}$, $\Delta f$ (and $k_{v0}$, $m_{v0}$) are recalled. 
The sizes of the resonators were inferred from SEM pictures (with a few \% resolution). The samples were symmetric within typically a few \% on the length $h$. Experimentally, the parameters $b_1$ to $b_3$ are found to be too small to be measured, and are thus considered to be zero.
We used tabulated values for the materials' mechanical properties$^($\footnote{We take for Silicon\cite{silicon} $E_z=160\,$GPa and $\rho=2.33\,$g/cm$^3$. For Aluminum\cite{paperthinAl}, we take $E_M=65\,$GPa and $\rho=2.70\,$g/cm$^3$. For NbTi\cite{handbook}, we use $E_M=80\,$GPa and $\rho=6.05\,$g/cm$^3$.}$^)$. The $\Delta f$ term is discussed in the next section. \\
The parameters $m_{v0}$ and $k_{v0}$ computed from theory are in very good agreement with the measurements. Also, the non-linear coefficients $b_1$ to $b_3$ are evaluated as being negligible, confirming the experiment. Indeed, for our extreme deflection (0.33$\,$mm$_{rms}$), the static distortion $X_0$ is smaller than a $\mu$m, the signal non-linearity amounts to about $-3\,$\% and the linewidth change is smaller than a mHz. We can thus safely state that the only geometrical non-linearity of concern is the $x^2$ resonance frequency dependence, through the parameter $b_0$. 
Experimentally, it is of order $0.1\,f_0/h^2$ (in Hz/m$_{rms}^2$ units).\\
The non-linear Rayleigh calculation of section \ref{nonlin} explains the sign of the $b_0$ parameter, the (small) decrease with the metal deposition seen in Fig. \ref{nonlinthick}, and the geometrical origin (that is $k_{v2}$ and $m_{v2}$ are dominant in the coefficient's expression). However, the magnitude of the effect fails to be predicted accurately by this simple formula. The coefficient calculated over-estimates the value fit on the data by a factor $3-5$, which is, considering the crude theoretical method, rather good.

\begin{table}[h!]
\begin{center}
\begin{tabular}{|c|c|c|}    \hline
Property                   &  Cb4           &    E6 (1$^{rst}$)         \\    \hline    \hline
 $l$                       &  $2.10\,$mm      & $1.55\,$mm                    \\    \hline
 $h$                       &  $1.85\,$mm      & $1.35\,$mm                    \\    \hline
 $a$                       &  $20\,\mu$m      & $27\,\mu$m                    \\    \hline
 $e'$                      &  $20\,\mu$m      & $32\,\mu$m                    \\    \hline
 $e$                       &  $4.5\,\mu$m     & $9.5\,\mu$m                   \\    \hline
 $\alpha$                  &  $0.5\,10^{-3}$  & $0.7\,10^{-3}$                \\    \hline    
$e_M$                      &  $150\,$nm       & $250\,$nm                \\    \hline    
metal used                 &  NbTi$^{(*)}$            & Al                    \\    \hline    \hline
$f_{res}$  at 4.2$\,$K     &  $1038.090\,$Hz  & $4194.665\,$Hz           \\    \hline
$\Delta f$ at 4.2$\,$K     &  $45\,$mHz       & $15\,$mHz                \\    \hline
spring $k_{v0}$ at 4.2$\,$K&  $0.029\,$N/m    & $0.95\,$N/m              \\    \hline
mass $m_{v0}$              &  $0.67\,\mu$g    & $1.35\,\mu$g             \\    \hline
$b_0$ experimental         &  $+0.45\,10^{8}\,$Hz/m$_{rms}^2$          & $+2.3\,10^{8}\,$Hz/m$_{rms}^2$        \\    \hline    \hline

$k_{v0}$ theoretical       &  $0.030\,$N/m          & $0.98\,$N/m         \\    \hline
$m_{v0}$ theoretical       &  $0.63\,\mu$g          & $1.36\,\mu$g        \\    \hline 
  
$b_{0}$ theoretical        &  $+1.65\,10^{8}\,$  Hz/m$_{rms}^2$   & $+12.7\,10^{8}\,$Hz/m$_{rms}^2$       \\    \hline   
$b_{1}$ theoretical        &  $-2.0\,10^{3}\,$   Hz/m$_{rms}^2$   & $-1.3\,10^{3}\,$Hz/m$_{rms}^2$        \\    \hline
$b_{2}$ theoretical        &  $+2.3\,$ m/m$_{rms}^2$              & $+7.6\,$m/m$_{rms}^2$                 \\    \hline
$b_{3}$ theoretical        &  $-1.8\,10^{5}\,$m$_{rms}^{-2}$      & $-3.2\,10^{5}\,$m$_{rms}^{-2}$        \\    \hline  
\end{tabular}
\caption{\label{NonlinSamples}Non-linear parameters and geometry of two typical samples studied in this article (experimental and theoretical). Spring constant and mass are deduced experimentally from the height of the peak $V'$ (given $\Delta f$) and the resonance frequency. $(*)$ NbTi is superconducting at 4.2$\,$K; See part \ref{last} for a discussion.}
\end{center}
\end{table}

\subsection{Metal properties}
\label{themetal}

The damping parameters $\Delta f$ and $\Lambda_{1,2}'$, together with the exact position of the resonance $f_{res}$ originate in materials' properties. In our configuration, we expect them to be dominated by the metallic layer. In order to unambiguously prove it, we deposited Aluminum three times on the same sample (E6).  
The first deposition was of 250$\,$nm on the top, the second added 260$\,$nm on top {\em but also} 500$\,$nm on the bottom, and the last deposition added 400$\,$nm more on the top.

\begin{figure}
\centerline{\includegraphics[width= 9. cm]{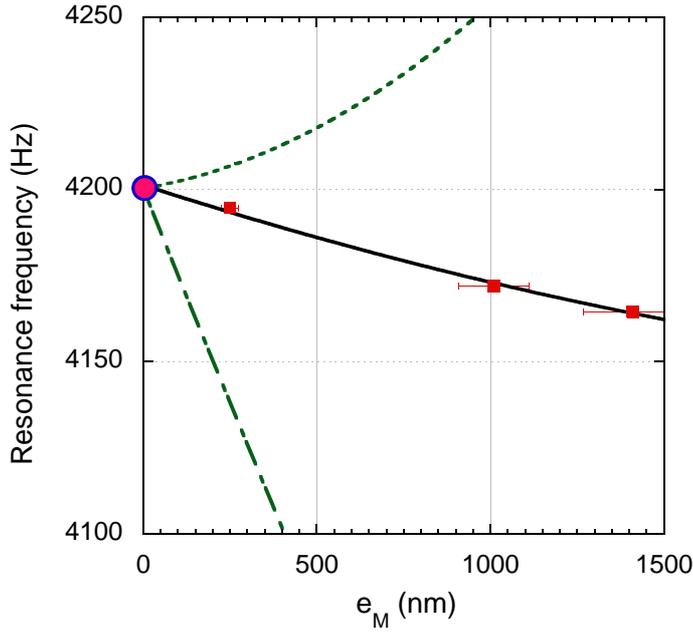}}
%
\caption{(Color online). Vacuum resonance frequency (in the linear regime, extrapolated at $T=0\,$K) as a function of the metal thickness. The large dot at zero thickness is an extrapolated value. The dot-dashed line is calculated while taking into account {\em only} the metallic mass. The dashed line takes also into account the finite Young's modulus $E_M$ of Aluminum. The full line is the result of the complete calculation, with a stress term $S$ (see text).}  
\label{f0mass}
\end{figure}

\vspace{0.15in}
In Fig. \ref{f0mass} we present the resonance frequency in vacuum extrapolated at 0$\,$K as a function of the metal thickness. The effect of temperature on $f_{res}$ and $\Delta f$ is postponed to the following paragraphs. The curves are calculated using the expressions$^($\footnote{The harmonic expressions, which do not include the $\alpha_0$ (thickness gradient) factor, have been scaled (within a small correction) on the results of the Rayleigh approach, both fitting the data. The effect of the thickness gradient on the determination of $S$ has been neglected.}$^)$ of section \ref{harmo}. The dot-dashed and dashed lines take into account the mass alone, and then the mass plus the elasticity of the metallic layer respectively. Both are unable to describe the measured data, although the effect is clearly linked to the metal. \\
We thus have to consider other mechanisms leading to frequency shifts. The first obvious one is {\em stress stored in the structure}. According to Eq. (\ref{mode}), a stressing force $S$ shifts downwards the resonance frequency $\omega_{res}$ by an amount $\delta \omega = \pm u^2 \omega_0\,\phi_0$ with $u=\sqrt{\frac{\left|S\right| h^2}{E_z I_z}}$ ($\phi_0$ tabulated, $u \ll 1$). The $\pm$ sign stands for tension ($S>0$) or compression ($S<0$). Adjusting the stress $S$ as a function of the thickness $e_M$ gives the thick line fit in Fig. \ref{f0mass}.\\
On the other hand, the shift could occur through a {\em reactive component of the friction mechanism}. According to section \ref{forced}, a reactive component $\Lambda_2$ shifts downwards the resonance frequency $\omega_{res}$ by an amount $\delta \omega = \omega_0\,\Lambda_2/m_{v0}$ (we identified $\omega$ to $\omega_0$, and the normal mass to  the vibrating mass).

\vspace{0.15in}
Clearly, a reactive component is {\em formally} equivalent to a stress term, through $\Lambda_2=\pm u^2\,m_{v0}\,\phi_0$. 
Can we distinguish between these two causes of frequency shifts? What are the microscopic mechanisms behind the stress stored in the structure, and the friction process?\\
The above questions are linked. We expect the effect of stored stress to be much greater than the reactive component, and almost temperature-independent in our temperature range (see below). Indeed, the frequency shifts shown in Fig. \ref{f0mass} for different thicknesses of metal are about 30 times greater than the shifts associated to temperature variations (Fig. \ref{frictionreactdissipe}). Moreover, the reactive shift {\em is} temperature-dependent. We thus claim that within a few \%, the black line fitting the data in Fig. \ref{f0mass} is solely due to the stress term. \\
Linking the bulk effective parameter $S$ to physical properties of the metal is rather difficult. In fact, the effect of the stress generated by a deposited layer on the resonance of the structure is a debated topic\cite{surfacestress,surfacestressChen,surfacestressMcF}.
In principle, a surface stress has no influence on the resonance of a cantilever beam, because it is balanced exactly by a corresponding bulk stress\cite{surfacestressGurtin}. However, this simple argument contradicts more recent studies\cite{surfacestressChen,surfacestressMcF}. Considering the two sides of one foot of the structure, the problem can be recast into a dependence to the total (tangential) stress due to the interfaces $\sigma_{total}=\sigma_{top}+\sigma_{bottom}$, and a dependence to the unbalance of the stresses $\Delta \sigma =\sigma_{top}-\sigma_{bottom}$. \\
\begin{figure}
\centerline{\includegraphics[width= 9. cm]{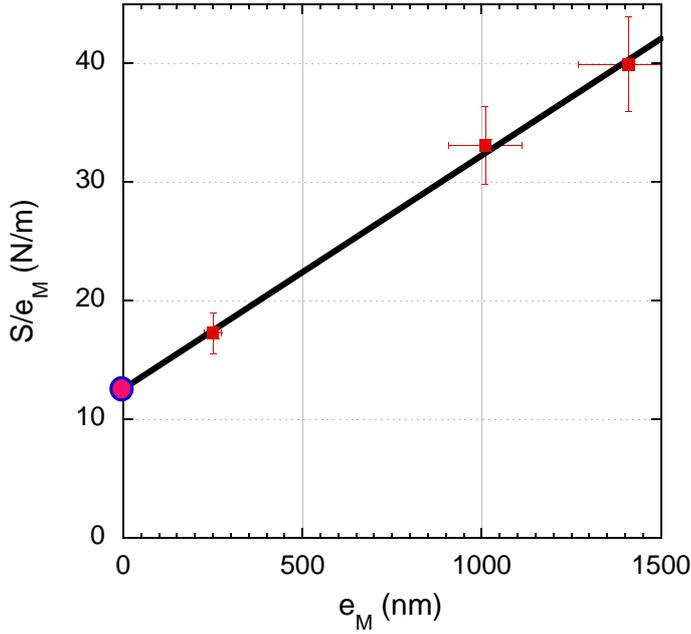}}
%
\caption{(Color online). Tensile stress ($S>0$) stored in the Silicon structure at low temperatures, normalized by the metal thickness $e_M$ (from fit Fig. \ref{f0mass}). The large dot at zero thickness is an extrapolated value. The line is a linear fit through the points.}  
\label{normS}
\end{figure}
In Fig. \ref{normS} we present the stressing force $S$ fitted in Fig. \ref{f0mass} divided by the total metal thickness $e_M$. The parameter $S/e_M$ is a smooth function of $e_M$, and no special anomaly can be seen on the second point (1010$\,$nm), where the balance is practically ideal ($\Delta \sigma \approx 0$, with about 500$\,$nm of metal deposited on both sides). We thus have to conclude that the effective parameter $S$ is linked solely to the total surface stress $\sigma_{total}$. \\
Comparing our expression Eq. (\ref{mode}) to a (rigorous) computation based on surface stresses\cite{surfacestressMcF}, we note that our simple model captures the correct behavior. The (remaining) stress stored in the Silicon $\sigma_{Si}=S/(e' e)$ will scale as $\sigma_{total}$ (the metal-induced surface stress), and for thin films would also scale\cite{microtribo} as $e_M/e$.
However, in Fig. \ref{normS} $S/e_M$ still exhibits a linear dependence to $e_M$. This suggests a proportionality between the total surface stress $\sigma_{total}$ and the {\em volume} of the deposited metallic layer. The non-zero value of $S/e_M$ in Fig. \ref{normS} at $e_M=0\,$nm would be a signature of the {\em ultimate residual stress} of the monocrystalline Silicon itself, generated by the matter/vacuum interface\cite{maugis,Haiss}.  \\
We extract as a maximum value $\sigma_{Si}=0.2\,$MPa which is rather low, compared to all the stressing sources involved in the problem. For instance, the mismatch between thermal expansion coefficients\cite{microtribo} would lead to much greater figures. The modeling presented here is thus clearly only indicative. Nonetheless the stress at the interface, due to thermal expansion or internal stresses (in the metal, like those related to grain boundary growth), has to be independent of temperature below 30$\,$K. Moreover, the stress generated by the distortion of the driven structure reaches 1$\,$GPa without destroying the metallic layer. This suggests that adhesion is perfect; no shear stresses are present at the interface and only tangential forces have to be considered\cite{microtribo}.

\begin{figure}
\centerline{\includegraphics[width= 9. cm]{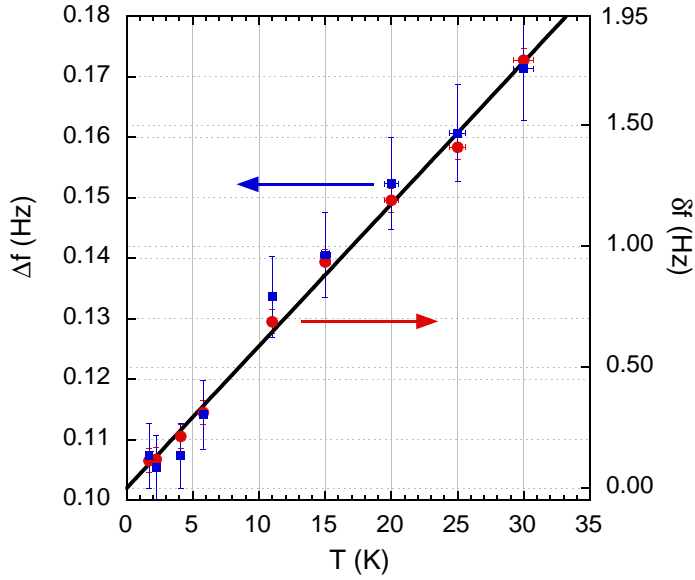}}
%
\caption{(Color online). Vacuum width $\Delta f$ and frequency shift $\delta f$ of the resonance, as a function of temperature (sample E6, third metal deposition). The full line is a linear fit.}  
\label{frictionreactdissipe}
\end{figure}

\begin{figure}
\centerline{\includegraphics[width= 9. cm]{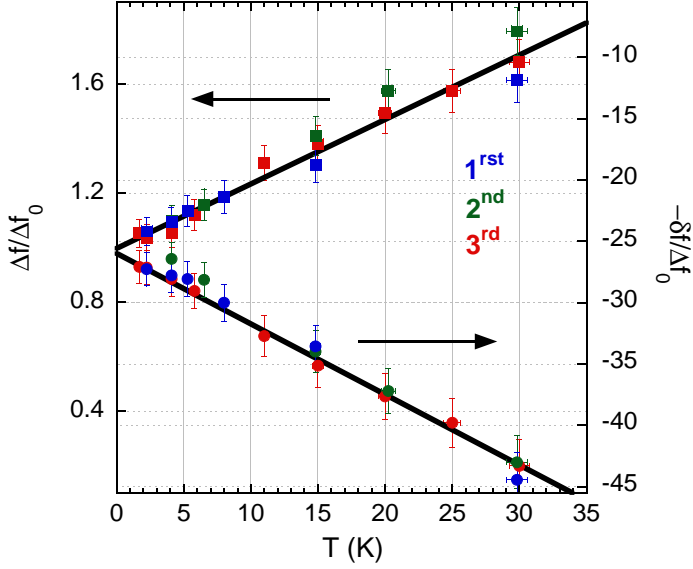}}
%
\caption{(Color online). Vacuum width $\Delta f$ and relative position $-\delta f$ of the resonance, as a function of temperature for the three runs on E6 with various metal quantities (Aluminum). For clarity, the frequency shift is presented with its negative sign. The data of each run has been normalized by the linewidth extrapolated at zero temperature $\Delta f_0$ (see text).}  
\label{frictionAll}
\end{figure}

\vspace{0.15in}
Measuring (in vacuum) the position and the linewidth of the resonance in the linear regime brings typically the results of Fig. \ref{frictionreactdissipe}, as a function of temperature. The (intrinsic) friction mechanism clearly produces dissipative and reactive components: the line shifts down, while it broadens when the temperature increases. Both terms are {\em linear} with respect to temperature in the range 1.5$\,$K to 30$\,$K. Since a shift occurs also through the stress force $S$, in order to be fully quantitative about the reactive component an assumption has to be made: in the following analysis we will consider that reactive and dissipative components are simply {\em proportional to each other}, a first order result obtained in the "weakly viscous" friction limit. \\
This corresponds in the data analysis to choosing a finite zero-temperature reactive shift for the friction.
As we increase the quantity of metal deposited, the damping (together with the reactive component) increases.
At 4.2$\,$K we get about 15$\,$mHz linewidth for the first metal deposition, and about 110$\,$mHz for the last. Both components remain linear with respect to temperature. Moreover, extracting for each metal deposition run an extrapolated value of the linewidth at zero temperature, say $\Delta f_0(e_M)$, we can normalize both the damping and the reactive component producing Fig. \ref{frictionAll}. 
As a result, {\em all data fall on the same linear dependence}, which may be empirically expressed as:
\begin{eqnarray*}
\Delta f (T) & = & \Delta f_0(e_M) \, (1+ \beta \, T ) , \\
\delta f (T) & = & \Delta f_0(e_M) \, \gamma \, (1+ \beta \, T )
\end{eqnarray*}
with $\beta= 0.024 \pm 0.003\,$K$^{-1}$ and $\gamma= 25.5 \pm 1$, independent of the metal quantity. These parameters have to be characteristic of the friction mechanism occurring inside the Aluminum layer.

\begin{figure}
\centerline{\includegraphics[width= 9. cm]{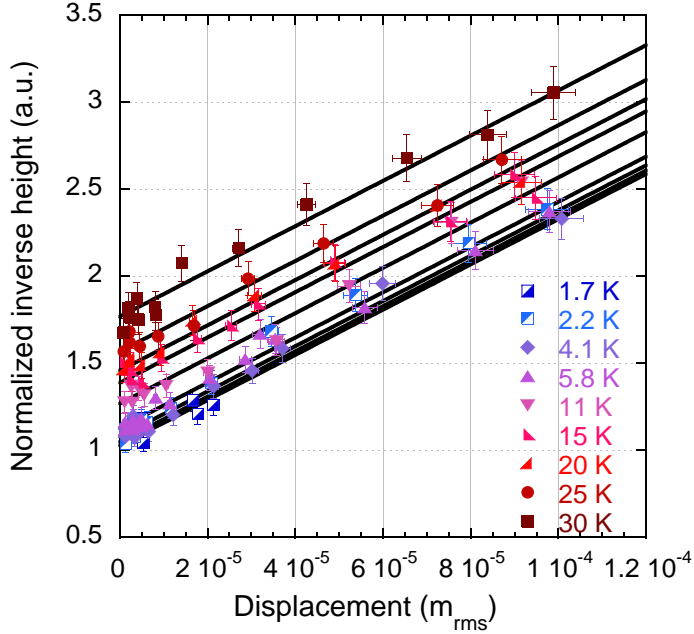}}
%
\caption{(Color online). Normalized inverse height $1/(V'_{max}/F_{ac})$ of the resonance peak measured in vacuum for sample E6 (third metal deposition), for various temperatures, as a function of displacement $X_{max}$. The 1 value on the $y$ axis is chosen for the $T=0\,$K extrapolated data at zero drive. The same linear dependence to the displacement (thick lines) is measured for all temperatures (see text). }  
\label{nonlinMetal}
\end{figure}

\vspace{0.15in}
From Fig. \ref{Kfact} we know that the damping process is also non-linear; we introduced in section \ref{dissipe} the parameters $\Lambda_{1,2}'$ to describe this effect. 
In Fig. \ref{nonlinMetal} we present the normalized inverse height $1/(V'_{max}/F_{ac})$ of the resonance curve measured as a function of the displacement for different temperatures. The same linear dependence as in Fig. \ref{Kfact} is seen, surprisingly with the {\em same slope} for all temperatures. 
We thus conclude that the $\Lambda_{1}'$ parameter is independent of temperature. \\
The non-linear signature on the reactive component (namely $\Lambda_{2}'$) is much more difficult to measure, and proper fits of the resonance lines are necessary. The shape of the frequency shift in Fig. \ref{tempnonlin} with its little minimum around $1.\,10^{-5}\,$m$_{rms}$ is actually due to the non-linear first order reactive shift which opposes the quadratic geometrical shift. As a result, $\Lambda_{2}'$ comes out to be also temperature independent. For the run on E6 with its first metal deposition, the non-linear term  $\Lambda_{2}'$ was too small to be seen. \\
Although $\Lambda_{1}'$ and $\Lambda_{2}'$ are temperature-independent, they strongly depend on the metal film thickness, which corroborates their material-dependent origin.

\begin{figure}
\centerline{\includegraphics[width= 9. cm]{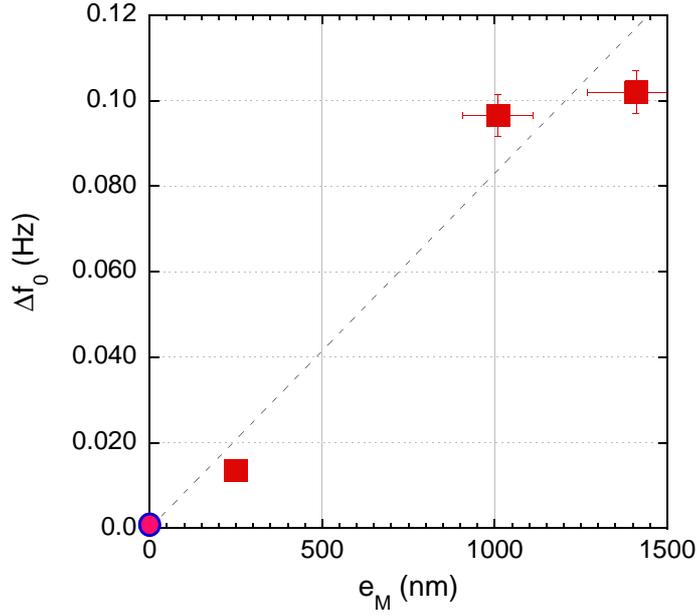}}
%
\caption{(Color online). Zero temperature intrinsic linewidth as a function of the metal thickness (sample E6, Aluminum). The large dot at zero thickness is an extrapolated value (which would neglect the Silicon itself). The dashed line is a guide to the eye. }  
\label{Df0}
\end{figure}

\begin{figure}
\centerline{\includegraphics[width= 13. cm]{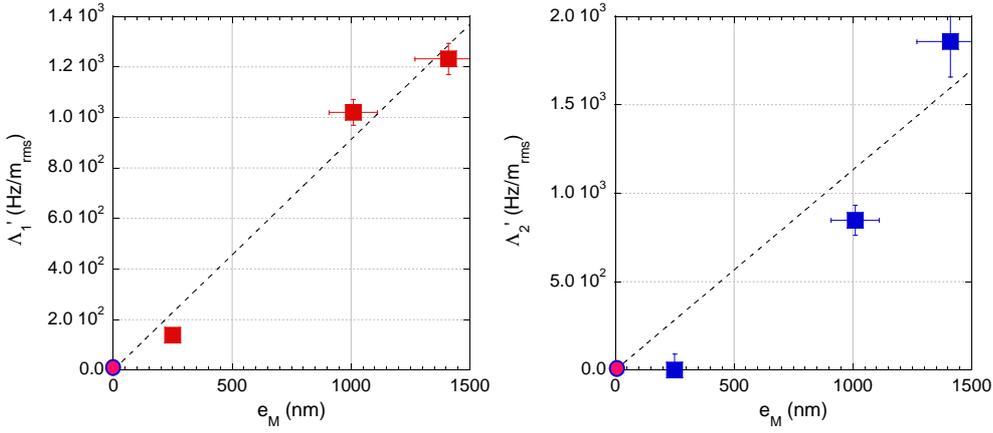}}
%
\caption{(Color online). Non-linear friction parameters $\Lambda_{1}'$ and $\Lambda_{2}'$ (sample E6, Aluminum). The large dots at zero thicknesses are extrapolated zeros, and the dashed lines are guides to the eye.}  
\label{Df0nonlin}
\end{figure}

\vspace{0.15in}
In order to conclude our experimental study of friction, we present in Fig. \ref{Df0} the zero temperature linewidth $\Delta f_0(e_M)$ as a function of the metal thickness deposited on sample E6. 
In Fig. \ref{Df0nonlin} the non-linear damping parameters ($\Lambda_{1}'(e_M)$ and $\Lambda_{2}'(e_M)$) are shown as a function of $e_M$. Both plots clearly demonstrate a strong variation with the metal thickness (or volume). They are characteristic of the friction mechanisms taking place in the metal layer$^($\footnote{The coefficients fit on the data could thus depend on frequency. The values discussed in the text stand for Aluminum around 4.2$\,$kHz (sample E6).}$^)$. \\
One dissipation mechanism of importance for small structures is {\em thermoelastic damping}\cite{theromelRoukes}.
In the MEMs language, it denominates the non-linear interaction between acoustic waves (such as the normal modes of the structure) and the bath of phonons (the expression of temperature). However, the {\em nature} of this non-linear interaction leading to friction could vary from one system/material to the other. 
Surprisingly, polycrystalline metal wires were found to display {\em amorphous-like} friction properties at low temperatures\cite{frictionmechanisms}. 
The nature of the friction process taking place in our devices could be debated.
One could speculate that the disorder present in the film due to the distribution of grain boundaries has to be involved. The properties discussed in this article would then be quite general, with variations in the numerical parameters depending on the metal under study. A summary of the dampings measured on various samples is given in Tab. \ref{damp-stat}.

\begin{table}[h!]
\begin{center}
\begin{tabular}{|c|c|c|c|c|}    \hline
Property                   &  Cb4           &  Eb2               &    E6 - 1$^{rst}$,2$^{nd}$, 3$^{rd}$      & Osaka\cite{japanese}   \\    \hline    \hline
$h$                        &  $1.85\,$mm    &   $1.58\,$mm       & $1.35\,$mm               & $2\,$mm  \\    \hline
$e'$                       &  $20\,\mu$m    &   $8.5\,\mu$m      & $32\,\mu$m               & $100\,\mu$m  \\    \hline
metal used                 &  NbTi$^{(*)}$  &  Al/Nb$^{(*)}$     & Al                       & Au  \\    \hline    
$e_M$                      &  $150\,$nm     &  $430\,$nm / $120\,$nm & $250-1010-1410\,$nm  & ? \\    \hline    \hline
$f_{res}$  at 4.2$\,$K     &  $1038\,$Hz    &  $4695\,$Hz & $4195-4172-4165\,$Hz          & $8700\,$Hz \\    \hline
$\Delta f$ at 4.2$\,$K     &  $45\,$mHz     &  $130\,$mHz & $15-105-110\,$mHz             & $75\,$mHz  \\    \hline    
\end{tabular}
\caption{\label{damp-stat}Summary of the dampings $\Delta f$ measured for various samples. Note the composite metal layer of sample Eb2. $(*)$ Metals superconducting at 4.2$\,$K; See part \ref{last} for a discussion.}
\end{center}
\end{table}

\subsection{Damping in $^4$He gas}
\label{4he}

A direct application of Silicon vibrating wires at low temperatures is viscosimetry in $^3$He/$^4$He gases or liquids. 
Moreover, MEMs (like AFM beams) at higher temperatures could be used as biological tools when immersed into fluids\cite{AFMHydro}. A careful characterization of the hydrodynamical behavior of model 
MEMs/fluids systems is thus strongly required. 

\vspace{0.15in}
This brought us to test one sample (E6 after the 2$^{nd}$ and 3$^{rd}$ Aluminum depositions, see preceding section) in $^4$He gas at 4.2$\,$K.
The cell was pumped and cooled down to 4.2 K in the usual way. No temperature regulation was used, and we systematically took a vacuum reference resonance line measurement. Then, $^4$He gas was slowly admitted to the cell. The pressure was monitored at room temperature on the top of the cryostat. We waited at least half an hour after each introduction to reach equilibrium.  
Thermomolecular pressure corrections\cite{chernyak} were estimated for our setup (capillary diameter of the order of 1$\,$mm), and found to be smaller than $2\,$\% even at our lowest pressures.

\vspace{0.15in}
In order to describe the gas, we apply the well-known kinetic theory\cite{kinetic}. It leads to the expressions for the mean free path and the (dynamic) viscosity:
\begin{eqnarray*}
l_{m\!f\!p} & = & \frac{m_{gas}}{\sqrt{2} \, \rho_{gas} \, v_m \, \sigma_0} ,\\
\nu         & = & \frac{1}{5} \, \rho_{gas} \, v_m \, l_{m\!f\!p}
\end{eqnarray*}
with $v_m=\sqrt{8 \, k_B T/( \pi \, m_{gas})}$ the mean velocity of the gas particles and $\rho_{gas}$ the mass density (at the given temperature and pressure). $m_{gas}$ is the mass of one gas particle, while $\sigma_0$ is the collision cross section.\\
The speed of sound $c_{sound}$, which has to be compared to the speed of the moving Silicon wire, is:
\begin{displaymath}
c_{sound}=\sqrt{\frac{\gamma\, P}{\rho_{gas}}}
\end{displaymath}
with $\gamma=C_p/C_v$. We evaluate these expressions for $^4$He gas at low temperatures using the Van der Waals equation of state and tabulated values\cite{VdW}. In the pressure/temperature $(P,T)$ range studied, the agreement with experimental data\cite{NIST} is excellent; the departure from the ideal gas behavior remains small.  \\
The slip correction, which is small but measurable, has been based on theoretical work for rarefied gases\cite{slippapergas}, giving the simple relationship $\xi=l_{m\!f\!p}$ (with $\xi$ the slip length).

\begin{figure}
\centerline{\includegraphics[width= 9. cm]{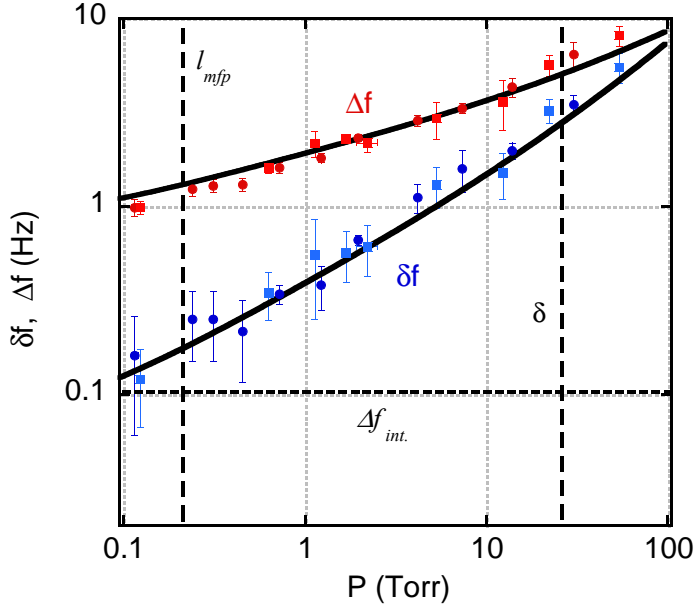}}
%
\caption{(Color online). Width $\Delta f$ and (down) shift $\delta \! f$ of the resonance, measured for small displacements in $^4$He gas at 4.21$\,$K. Squares and circles correspond to two runs with sample E6 covered by two slightly different metal thicknesses. Vertical lines correspond to the pressure below which $l_{m\!f\!p}$ becomes larger than $r$ (left), and the pressure above which $\delta$ becomes smaller than $r$ (right). The horizontal dashed line indicates the intrinsic damping magnitude, which has been subtracted. Full lines are fits explained in the text, without adjustable parameters.}  
\label{pressure}
\end{figure}

\vspace{0.15in}
The measured additional damping $\Delta f$ and frequency shift $\delta \! f$ of the structure's resonance (first mode), as a function of pressure, are presented in Fig. \ref{pressure}. The full lines are based on the theoretical calculations of section \ref{gas}, using the geometrical parameters of the wire, and {\em no free parameters}.\\
The speed $\omega\,x$ of the oscillating structure was kept below 1$\,$m/s (displacement $x$ smaller than $30\,\mu$m). 
Since the speed of sound $c_{sound}$ is about 100$\,$m/s, the "incompressibility" assumption is well satisfied. 
For the smallest drives used, the Reynolds number (defined as $(\omega \, x) (2 r) \rho_{gas} / \nu$) was always smaller than 1, and the displacement $x$ much smaller than the width $2 r$ of the structure. Making systematic measurements as a function of speed/displacement allowed us to experimentally define the linear regime of the damping force.
Moreover, in the whole pressure range $l_{m\!f\!p}/\delta \ll 1$ which guarantees that hydrodynamics apply. \\
Since $\delta/l \ll 1$ and $\delta/h \ll 1$ the two feet and the paddle can be treated as infinite, applying the Rayleigh method to take into account the variation of the speed along each foot.\\
However, the agreement on Fig. \ref{pressure} seems to get worse as $l_{m\!f\!p}$ becomes larger than $r$, and $\delta$ smaller than r (with $r$ the wire's half width). On the low pressure side ($l_{m\!f\!p} \geq r$) the slip correction seem to be smaller than measured. However, the deviation could be due to the finite size of the experiment, since the closest wall to the wire (the Silicon chip itself) is only 300$\,\mu$m away (a value reached by $\delta$ around 0.1$\,$Torr). On the high pressure side ($\delta \leq r$), one could argue that the thickness $e$ of the structure starts to play a role, and the correction function $\Omega$ valid for $e \ll 2r$ (section \ref{gas}) would have to be reevaluated. \\
Nonetheless, the agreement is fairly good (within the experimental error bars arising from the smallness of the signals), without adjustable parameters.

\section{A BIT BEYOND}
\label{last}

In this section we present some features of our mechanical devices which, while mostly {\em outside} of the scope of this paper, help defining the applicability range of our results. \\
In a first section, very strong drive features (in vacuum at 4.2$\,$K) are discussed, and how they can be related to our description.
The second section comments on measurements realized below 1.5$\,$K in vacuum. 

\subsection{Extreme drive effects}

\begin{figure}
\centerline{\includegraphics[width= 12. cm]{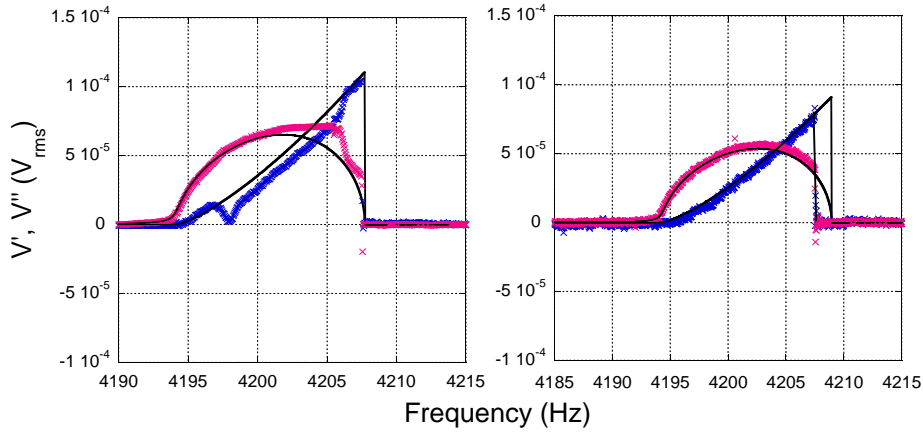}}
%
\caption{(Color online). Anomalies observed on the (first mode) resonance of sample E6 (with its first metal deposition) at very strong drives. Measurements are done at 4.2$\,$K in vacuum using magnetic fields of the order of 10$\,$mT. Left: a multi-peaked structure has appeared. Right: the resonance "switches off$\,$" earlier than expected. Both lines were measured with a force of the order of 3$\,$nN$_{rms}$, corresponding to displacements of the order of 0.2$\,$mm$_{rms}$ (speeds 6$\,$m/s $\!_{rms}$). The thick lines are fits obtained from part \ref{datanalys}.  }  
\label{toostrong}
\end{figure}

In Fig. \ref{toostrong} we reproduce two resonance lines observed at very strong drives. Typically, above displacements of the order of 0.15 mm$_{rms}$ (for a structure with width 10$\,\mu$m and length 1$\,$mm approximately), irreproducible irregularities may appear. \\
One type of irregularities is the appearance of an additional structure on the resonance lines (left, Fig. \ref{toostrong}). The reason of the distortion is not yet known, but one could argue that mode couplings could be the cause. Effectively, the stress stored in the material together with all non-linear terms involved in the dynamics equation, do couple the various modes of vibrations of the structure and their harmonics. The theory presented in part \ref{theor} neglected these effects, and the amplitudes of these higher modes were always taken to be zero. The suggestion would be that, sometimes, fluctuations in the background noise redistribute the amplitude weight in a non-negligible way toward higher modes. \\
The second type of irregularities observed is an early return to stillness (right, Fig. \ref{toostrong}).
Again, the exact reason for this behavior is not known. Since two excitation branches are stable in this domain of frequencies (observed usually by sweeping up, or down, the resonance line), one can speculate that a fluctuation of some sort may make the system switch from one state to the other\cite{bistablePRB} (bistability). However, note the abrupt return to zero voltage on the quadrature signal, which draws usually (in most of the measurements) a quasi-circular shape (corresponding to the black theoretical fit). 
It is thus quite easy to distinguish "irregular" measurements from normal ones, an attitude we adopted in part \ref{datanalys}.

\begin{figure}
\centerline{\includegraphics[width= 9. cm]{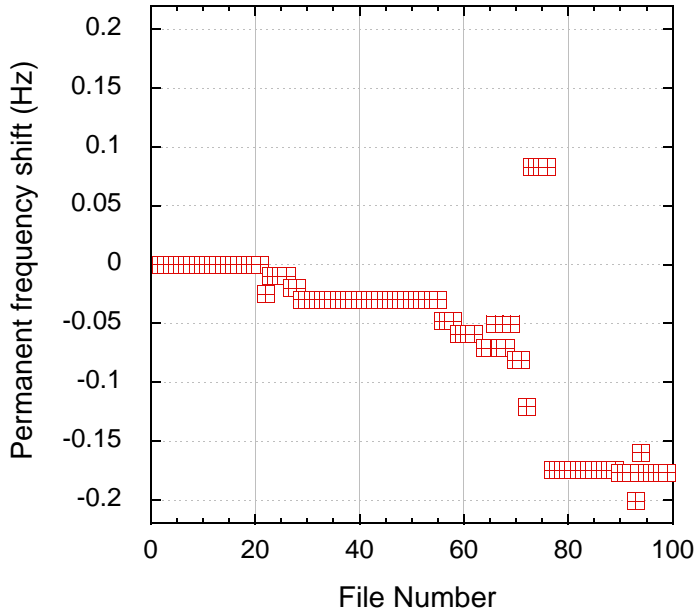}}
%
\caption{(Color online). Shift of the resonance position of sample E6 (with its first metal deposition), after many high-drive measurements (taken at 4.2$\,$K in vacuum) and thermal cyclings. The $y$ axis is the position of the resonance down from the reference value 4194.665$\,$Hz. The $x$ axis, which corresponds to the number of the $n^{th}$ file, can be viewed as a count of the high-drive/small-drive cyclings and/or temperature cyclings (see text).  }  
\label{aging}
\end{figure}

\vspace{0.15in}
As we cycle the mechanical device from high-drives to low-drives, and from high temperatures to low temperatures, we note that  the {\em resonance frequency shifts permanently} (Fig. \ref{aging}). \\
For the first 70 files, the driving force was kept below 200$\,$pN (displacement smaller than 50$\,\mu$m$_{rms}$). During this period, many thermal cyclings occurred, most of them between 4.2$\,$K and temperatures of the order of 50$-$70$\,$K, and a few up to 300$\,$K. The absolute position of the resonance (in the linear regime) shifted by less than $\pm$25$\,$mHz from one set to the other. The figure is remarkable$^($\footnote{The stability certainly benefits from the goalpost-shape of the structure, as opposed to a simple doubly-clamped beam which potentially stores a substantial stress at low temperatures.}$^)$, since it corresponds to a stability of the order of $\pm$6$\,$ppm. \\
After the 70$^{th}$ file, we started to use high drives (typically displacements from 100$\,\mu$m$_{rms}$ to 0.3$\,$mm$_{rms}$). The stability gets much worse, and the permanent shifts observed are much bigger. Quoting numbers is rather difficult, due to the irreproducibility of the phenomenon. The permanent shifts even seemed sometimes to settle or appear {\em after an additional thermal cycling}. These permanent shifts could then be as large as 0.2$\,$Hz. \\
At last, adding more metal on sample E6 seemed to make things worse. The "low drive" thermal cyclings seemed to have a stability of the order of $\pm$300$\,$mHz, and the high drive shifts could be as large as 1$\,$Hz.

\vspace{0.15in}
The exact mechanism behind these shifts can be debated. We propose that they arise from the {\em plasticity of the (stressed) metal layer} (at strong drives and/or after thermal cyclings). Indeed, we certainly go beyond the elastic yield limit of Aluminum for our strongest excitations.  The idea is that the distortion of the metal layer changes the stress $S$ stored (defining the parameter $u$ introduced in part \ref{theor}), especially after a "high temperature" annealing of the sample.
The permanent shift has a global trend toward {\em lower} frequencies, which cannot be explained by a hardening of the metal's Young modulus $E_M$. On the other hand, in part \ref{datanalys} we claimed that the stored stress made the resonance frequency decrease, which is in agreement with this interpretation.

\subsection{Below 1.5$\,$K}

\begin{figure}
\centerline{\includegraphics[width= 8. cm]{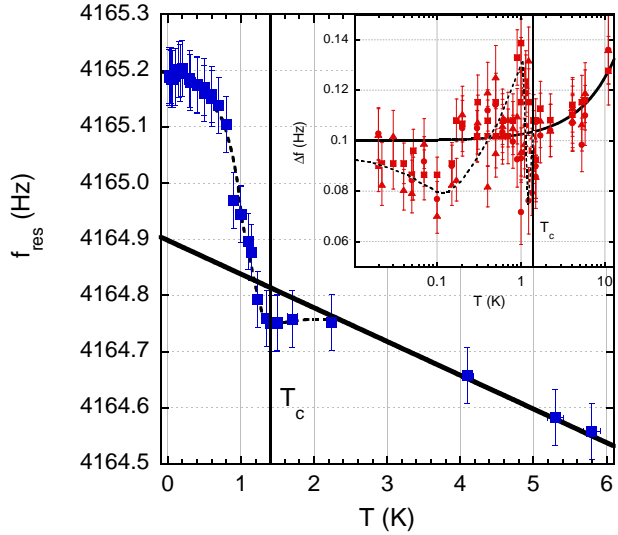}}
%
\caption{(Color online). Signature on the friction mechanism of the thin Aluminum film superconductivity (sample E6, third metal deposition). The $T_C$ is marked by vertical lines. The main graph presents the frequency shift of the resonance line toward higher values (reactive effect). The inset presents the measured linewidth (symbols are for different measurement techniques), which seem to decrease (dissipative component; note the logarithmic temperature scale). The full lines are the linear fits presented in parts \ref{experiment} and \ref{datanalys}. The dashed lines are guides to the eyes.}  
\label{supra}
\end{figure}

\begin{figure}
\centerline{\includegraphics[width= 8. cm]{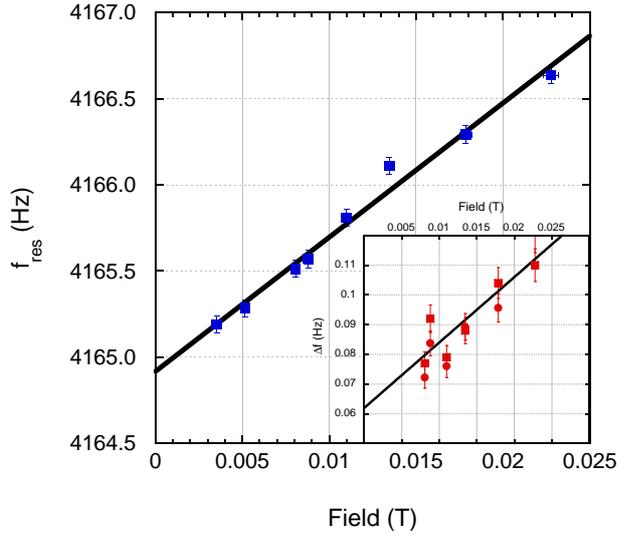}}
%
\caption{(Color online). Effect of the magnetic field $B$ on the metal-dependent parameters $f_{res}$ (resonance frequency, main graph) and $\Delta f$ (linewidth, inset) measured at 12 mK (sample E6, third metal deposition). The different symbols used for the linewidth plot correspond to various criteria. Linear fits extrapolate to the "zero field limit" (see discussion in the text).}  
\label{fieldeffect}
\end{figure}

This article focuses on {\em normal metal-covered} structures. For our resonators we used Aluminum, Niobium, or Niobium-Titanium which are superconducting below a given temperature $T_C$. \\
We present briefly the features observed for one sample (Aluminum covered) studied in vacuum down to 10$\,$mK. 
The key point is that the analytical fit procedure presented in this article is still valid, {\em but the material-dependent parameters we introduced change behavior} below $T_C$ (both linear $f_{res}$, $\Delta f$ and non-linear terms  as  $\Lambda_{1,2}'$). Indeed, on Fig. \ref{supra} we show the (linear) resonance frequency $f_{res}$ and the (linear intrinsic) damping parameter $\Delta f$ as a function of temperature. The frequency is found to increase suddenly, while the damping decreases (after a spiky region).
For the damping, different criteria are used on the data to extract the width (fit of in-phase, fit of out-of-phase, direct measure in linear regime, or measure of the inverse height). One cool-down is represented, for a field of about 10$\,$mT.\\
On Fig. \ref{fieldeffect}, for one particular cool-down we measured the field-dependence of these parameters around 12$\,$mK. Keeping the drive around 200$\,$pN$_{rms}$ (about \linebreak 7$\,\mu$m$_{rms}$ displacement) allowed us to preserve linearity. Again, the width is extracted using various criteria. Both $f_{res}$ and $\Delta f$ display a {\em linear} field dependence. 
The resonance frequency extrapolated at zero field corresponds well to the zero temperature value of the high temperature fit. The zero field width is on the other hand about 30$\,$\% smaller than the high temperature extrapolation at $T=0\,$K. 

\vspace{0.15in}
These effects are clearly linked to the superconductivity of the metallic layer. 
The basic idea is that by expelling the field lines (Meissner effect) the metallic layer feels a new "spring" force$^($\footnote{This force can even become non-linear at strong drives, which would imply a material-dependent $b_0$ parameter.}$^)$, which depends on the magnetic field. For a type I superconductor (perfectly diamagnetic), the magnetic energy is simply proportional to the length of the distorted field lines, and to their number. Due to the demagnetization factor, on the edges of the vibrating layer the field penetrates by normal conducting domains. These domains do not participate to the restoring force, but since the normal metal dissipation is larger than for superconducting ones, they increase the energy relaxation. \\
For a type II superconductor in the mixed state\cite{esquinazi}, flux lines penetrate the metal and vortices (pinned, or unpinned) participate to the restoring force. The movement of the vortices also generates dissipation, seen in the linewidth parameter. \\
The data presented in Figs. \ref{supra} and \ref{fieldeffect} concern a sample covered by a 1.4$\,\mu$m layer of Aluminum (type I superconductor). The $T_C$ of the layer, as seen in Fig. \ref{supra} is about 1.4$\,$K (coherence length $\xi_0 \approx 1.3\,\mu$m comparable to the thickness\cite{alusupra}) but the transition is rather broad, which proves that the metal is in its "dirty" limit (which was expected from the metal deposition technique, part \ref{experiment}).
In our data, no hysteretic effect could be observed.
However, eventhough all the discussed features are reproducible, their {\em quantitative magnitude seems to depend on each cool-down} (although the zero field parameters seem reproducible).
These effects clearly deserve more studies. In fact, torsional Silicon oscillators were already used as a tool to characterize superconducting materials\cite{bishopsupra}. In this view, our devices could prove to be especially useful.

\section{CONCLUSION}

In this article, we studied experimentally and theoretically the resonance of the first natural mode of Silicon vibrating wires. The structures were about  1$\,$mm wide with thicknesses in the micron range, and covered by metal layers. \\
Based on simple analytical models, we prove our ability to describe the dynamics of these objects up to very high drives, when the resonance is highly non-linear. We prove that the most striking non-linear effect comes from the geometry, and generates a quadratic dependence of the frequency shift to the displacement. The non-linear coefficient can be 
be calculated analytically by means of an extended Rayleigh method, with the correct properties and the right order of magnitude. The linear properties fit the model within the experimental precision. \\
The damping arises from the metal layers, and can be fully characterized. Both dissipative and reactive components are seen, which are simply proportional to each other and linear with temperature. As a consequence of the metal deposition, stress is stored in the structure which shifts the resonance frequency. The friction mechanism occurring inside the metal displays a non-linear behavior which is described by a simple linear relation of the reactive and dissipative components to the displacement. \\
When the metallic layers become superconducting, signatures can be observed on the resonance line, like in torsional oscillator studies\cite{bishopsupra}. Both the linewidth and the resonance frequency of the line are affected by superconductivity.
The damping experienced by the structure when immersed in $^4$He gas at 4.2$\,$K is fit to theory without free parameters. 

\vspace{0.1in}
Opening this work onto future experiments, we claim that these devices are promising for low temperature studies. They could be indeed used as tools for the characterization of metallic/superconducting thin films. The Silicon itself could also be studied (with less metal) and the results compared to existing data obtained with other geometries\cite{ParpiaQ}. \\
The goal-post Silicon shape is well suited for quantum fluid experiments. 
Many applications can be envisaged in $^4$He, $^3$He. As one of them, the ULTIMA\cite{ultima} project relies on the fabrication of ultra-sensitive superfluid $^3$He bolometers, which can be achieved through the Silicon technique. These devices could become standards for ultra-low temperature (vibrating wire) thermometry. 

\vspace{0.1in}
The physics of oscillating bodies immersed in quantum fluids, or studied at low temperatures in vacuum, has a long and rich history\cite{kleimanbishop,jeevak,CHH,pobellnonlin}. More recently, quartz tuning forks were proposed and studied as (cheap and easy-to-use) sensors in cryogenic environments\cite{quartzforkBuu,quartzforkRob}.
The internal damping of Silicon structures is much smaller, and their properties are much more stable and reproducible; the hydrodynamic regimes probed by these two techniques are also different. The versatility of Silicon microfabrication opens up a wide area of possibilities for low temperature techniques, for which our theoretical description is a tool enabling optimization. 

\vspace{0.1in}
{\bf Remark:} misprints quoted in Erratum {\it J. of Low Temp. Phys.} Volume {\bf 157}, Issue 5, Page 566 (2009) have been corrected in this version of the manuscript. 

\begin{acknowledgements}
This research was supported through the ULTIMA grant of the ``Agence
Nationale de la Recherche" (ANR), France (NTO5-2\_41909). The authors wish to
thank J. Parpia and S. Hentz for helpful discussions.
\end{acknowledgements}

\pagebreak


\begin{thebibliography}{99}

\bibitem{viscosity1} M.A. Black, H.E. Hall and K. Thompson, {\it J. Phys. C: Solid St. Phys.} {\bf 4}, 129 (1971).
\bibitem{viscosity2} J.M. Goodwin, {\it J. Phys. E: Sci. Instrum.} {\bf 6}, 452-456 (1973).
\bibitem{pickett} A.M. Gu\'enault, V. Keith, C.J. Kennedy, S.G. Mussett and G.R. Pickett, {\it J. of Low Temp. Phys.} {\bf 62}, 511 (1986).
\bibitem{MEMsBOOKS} J.A. Pelesko and D.H. Bernstein, {\it Modeling MEMs and NEMs}, Chapman \& Hall/CRC (2003).
\bibitem{heliumMEMs} A. Kraus, A. Erbe and R.H. Blick {\it Nanotechnology} {\bf 11}, 165-168, (2000).
\bibitem{ultima} C.B. Winkelmann, J. Elbs, E. Collin, Yu.M. Bunkov and H. Godfrin, {\it Nuclear Instruments and Methods in Physics Research, section A} {\bf 559}, 384-386 (2006).

\bibitem{kleimanbishop} R.N. Kleiman, G. Agnolet and D.J. Bishop, {\it Phys. Rev. Lett.} {\bf 59}, 2079 (1987).
\bibitem{jeevak} R.E. Mihailovich and J.M. Parpia, {\it Physica B} {\bf 165-166}, 125-126 (1990).

\bibitem{usPhysicaB} S. Triqueneaux, E. Collin, D.J. Cousins, T. Fournier, C. B\"auerle, Yu.M. Bunkov and H. Godfrin, {\it Physica B} {\bf 284}, 2141-2142 (2000).
\bibitem{japanese} Y. Hayashi, H. Nakagawa, H. Yano, O. Ishikawa and T. Hata, {\it Physica B} {\bf 329-333}, 108-109 (2003). 

\bibitem{CHH} D.C. Carless, H.E. Hall and J.R. Hook, {\it J. of Low Temp. Phys.} {\bf 50}, 583 (1983).

\bibitem{timoRDM} S. Timoshenko, {\it Theory of structures}, 2d ed. McGraw-Hill (1965).
\bibitem{timoVIBR} S. Timoshenko, D.H. Young and W. Weaver Jr.,  {\it Vibration problems in engineering}, fifth ed. John Wiley \& Sons (1990).

\bibitem{masspaper} Y.M. Tseytlin,  {\it Review of Scientific Instruments} {\bf  76}, 115101 (2005).
\bibitem{surfacestress} P. Lu, H.P. Lee, C. Lu and S.J. O'Shea, {\it Physical Review B} {\bf 72}, 085405 (2005).

\bibitem{silicon} Ed. by W.S. Trimmer, {\it Micromechanics and MEMs}, IEEE press (1997).
\bibitem{paperthinAl} M.A. Haque and M.T.A. Saif,  {\it Proceedings of the National Academy of Sciences of the United States of America} {\bf 101}, 6335–6340 (2004).
\bibitem{handbook} Ed. by R.C. Weast and M.J. Astle, {\it Handbook of Chemistry and Physics}, 63$^{rd}$ Edition, CRC Press (1983).
\bibitem{nonlinMEM} V. Kaajakari, T. Mattila, A. Oja and H. Sepp\"a, {\it Journal of Microelectromechanical Systems} {\bf 13}, 715 (2004).

\bibitem{dry} L.F.C. Zonetti, A.S.S. Camargo, J. Sartori, D.F. de Sousa, and L.A.O. Nunes, {\it Eur. J. Phys.} {\bf 20}, 85–88 (1999). 
\bibitem{Qfact} P. Mohanty, D.A. Harrington, K.L. Ekinci, Y.T. Yang, M.J. Murphy and M.L. Roukes, {\it Physical Review B} {\bf 66}, 085416 (2002).
\bibitem{ParpiaQ} R.E. Mihailovich and J.M. Parpia, {\it Phys. Rev. Lett.} {\bf 68}, 3052 (1992).
\bibitem{ParpiaQfield} R.D. Biggar and J.M. Parpia, {\it Physical Review B} {\bf 56}, 13638 (1997). 

\bibitem{stokes} G. G. Stokes, {\it Mathematical and Physical Papers} {\bf 3}, 38-54, London : Cambridge University Press (1901).
\bibitem{landaufluid} L.D. Landau and E.M. Lifshitz, {\it Fluid Mechanics},  second ed. Butterworth-Heinemann (1987).

\bibitem{Sader} J.E. Sader, {\it Journal of Applied Physics} {\bf 84}, 64 (1998).
\bibitem{landaumeca}  L.D. Landau and E.M. Lifshitz, {\it Mechanics}, third ed. Elsevier Science Ltd. (1976).

\bibitem{alu} A.L. Woodcraft, {\it Cryogenics}, {\bf 45}, 626-636 (2005).
\bibitem{alu2} R.L. Greene, C.N. King and R.B. Zubeck, {\it Physical Review B} {\bf 6}, 3297 (1972).

\bibitem{pobellnonlin} R. K\"onig, P. Esquinazi and F. Pobell, {\it J. of Low Temp. Phys.} {\bf 90}, 55 (1993).

\bibitem{microtribo} Ed. by B. Bhushan, {\it Handbook of Micro/Nano Tribology}, second ed. CRC Press (1999).
\bibitem{surfacestressGurtin} M.E. Gurtin, X. Markenscoff and R.N. Thurston, {\it Applied Physics Letters} {\bf 29}, 529 (1976).
\bibitem{surfacestressChen} G.Y. Chen, T. Thundat, E.A. Wachter and R.J. Warmack, {\it J. Appl. Phys.} {\bf 77}, 3618 (1995).
\bibitem{surfacestressMcF} A.W. McFarland, M.A. Poggi, M.J. Doyle, L.A. Bottomley and J.S. Colton, {\it Applied Physics Letters} {\bf 87}, 053505 (2005).

\bibitem{maugis} D. Maugis, {\it Contact, Adhesion and Rupture of Elastic Solids}, Springer (1991).
\bibitem{Haiss} W. Haiss, {\it Rep. Prog. Phys.} {\bf 64}, 591-648 (2001).

\bibitem{theromelRoukes} R. Lifshitz and M.L. Roukes, {\it Phys. Rev. B} {\bf 61}, 5600 (2000).
\bibitem{frictionmechanisms} P. Esquinazi, R. K\"onig and F. Pobell, {\it Z. Phys. B - Condensed Matter} {\bf 87}, 305-321 (1992).

\bibitem{pariaslip} D. Einzel and J.M. Parpia, {\it  J. of Low Temp. Phys.} {\bf 109}, 1 (1997).
\bibitem{bowley} R.M. Bowley and J.R. Owers-Bradley, {\it   J. of Low Temp. Phys.} {\bf 136}, 15 (2004).

\bibitem{AFMHydro} A. Maali, C. Hurth, R. Boisgard, C. Jai, T. Cohen-Bouhacina and J.-P. Aimé, {\it Journal of Applied Physics} {\bf 97}, 074907 (2005).

\bibitem{kinetic} F. Reif, {\it Fundamentals of Statistical and Thermal Physics}, McGraw-Hill (1965).
\bibitem{VdW} We use the Van der Waals equation of state $n R T = ( P + a n^2)(1 - b n)$ and the kinetic theory of gases with $^4$He tabulated values :  $a=3.46\,10^{-3}\,$Pa$($m$^3$/mol$)^2$, $b= 23.71\,10^{-6}\,$m$^3$/mol and $m_{gas}=6.64648\,10^{-27}\,$kg, $\sigma_0(4.21\,$K$)=1.1\,10^{-19}\,$m$^2$.
\bibitem{NIST} V.D. Arp and R.D. McCarty, {\it NIST TECHNICAL NOTE 1334, Thermophysical Properties of Helium-4 From 0.8 to 1500 K With Pressures to 2000 MPa} (1989).
\bibitem{slippapergas} G.H. Tang, W.Q. Tao and Y.L. He, {\it Physical Review E} {\bf  72}, 056301 (2005).

\bibitem{chernyak} V.G. Chernyak, B.T. Porodnov and P.E. Suetin, {\it Journal of Engineering Physics and Thermophysics} {\bf 26}, 309 (1974).

\bibitem{bistablePRB} C. Stambaugh and H.B. Chan, {\it Physical Review B} {\bf  73}, 172302 (2006). 

\bibitem{alusupra} J. Romijn, T.M. Klapwijk, M.J. Renne and J.E. Mooij,  {\it Physical Review B} {\bf  26},  3648 (1982). 
\bibitem{esquinazi} P. Esquinazi, {\it   J. of Low Temp. Phys.} {\bf 85}, 139 (1991).

\bibitem{bishopsupra} P.L. Gammel, A.F. Hebard and D.J. Bishop, {\it Phys. Rev. Lett.} {\bf 60}, 144 (1988); P.L. Gammel, L.F. Schneemeyer, J.V. Waszczak and D.J. Bishop, {\it Phys. Rev. Lett.} {\bf 61}, 1666 (1988).

\bibitem{quartzforkBuu} D.O. Clubb, O.V.L. Buu, R.M. Bowley, R. Nyman and J.R. Owers-Bradley, {\it  J. of Low Temp. Phys.} {\bf 136}, 1 (2004).
\bibitem{quartzforkRob} R. Blaauwgeers, M. Blazkova, M. \v{C}love\v{c}ko, V.B. Eltsov,
R. de Graaf, J. Hosio, M. Krusius, D. Schmoranzer, W. Schoepe,
L. Skrbek, P. Skyba, R.E. Solntsev and D.E. Zmeev, {\it  J. of Low Temp. Phys.} {\bf 146}, 537 (2007).

\end{thebibliography}
\end{document}